\documentclass[10pt,journal,compsoc]{IEEEtran}

\usepackage[ruled,linesnumbered,lined,noend]{algorithm2e}
\usepackage{lineno}
\usepackage{cite}
\usepackage{graphicx}
\usepackage{subfigure}
\usepackage{pifont} 
\usepackage{url}
\usepackage{color}
\usepackage{booktabs}
\usepackage{multirow}
\usepackage{latexsym,bm,amsmath,amssymb}
\usepackage{array}
\usepackage{algpseudocode}
\usepackage{amsmath}
\usepackage{lscape}
\usepackage{comment}
\usepackage{balance}
\usepackage[colorlinks,linkcolor = blue,citecolor = blue,citecolor = blue,filecolor = blue,urlcolor=blue]{hyperref}
\usepackage{tcolorbox}
\usepackage{ragged2e}
\usepackage{multirow}
\usepackage{longtable}
\usepackage{ragged2e}
\usepackage{array}
\usepackage{xltabular}
\usepackage{booktabs}
\usepackage{multirow}
\usepackage{xtab} % 添加在导言区
\usepackage{array}
\usepackage{multirow}
\usepackage{booktabs}
\usepackage{booktabs}
\usepackage{longtable}
\usepackage{array}
\usepackage{booktabs}
\usepackage{multirow}
\usepackage{stfloats}
\usepackage{lscape} % Optional: for landscape orientation if needed
\usepackage{ragged2e} % For better control over text alignment
\definecolor{light-gray}{gray}{0.95}

\definecolor{light-gray}{gray}{0.85}

\definecolor{mygray}{gray}{.9}
\begin{document}
\title{An Empirical Study of OpenAI API Discussions on Stack Overflow}
%候选标题

\author{Xiang Chen,
        Jibin Wang,
        Chaoyang Gao,
        Xiaolin Ju,
        Zhanqi Cui
        
\IEEEcompsocitemizethanks{

\IEEEcompsocthanksitem Xiang Chen with School of Artificial Intelligence and Computer Science, Nantong University, China, and also with State Key Lab. for Novel Software Technology, Nanjing University, Nanjing, China. E-mail: xchencs@ntu.edu.cn

Jibin Wang with School of Artificial Intelligence and Computer Science, Nantong University, China, and also with Chang Chien College, Nantong University, China
E-mail: W20040830@outlook.com

Chaoyang Gao and Xiaolin Ju are with School of Artificial Intelligence and Computer Science, Nantong University, China.
E-mail: gcyol@outlook.com, ju.xl@ntu.edu.cn

Zhanqi Cui with Computer School, Beijing Information Science and Technology University.
E-mail: czq@bistu.edu.cn

\IEEEcompsocthanksitem Xiang Chen and Jibin Wang contributed equally to this work and are recognized as co-first authors.

\IEEEcompsocthanksitem Xiang Chen is the corresponding author.

}% <-this % stops an unwanted space
\thanks{Manuscript received April 19, 2020; revised August xx, xxxx.}}

\markboth{IEEE Transactions on Software Engineering,~Vol.~14, No.~8, August~2015}%
{Shell \MakeLowercase{\textit{et al.}}: Bare Demo of IEEEtran.cls for Computer Society Journals}

\IEEEtitleabstractindextext{
\begin{abstract}
The rapid advancement of large language models (LLMs), represented by OpenAI's GPT series, has significantly impacted various domains such as natural language processing, software development, education, healthcare, finance, and scientific research.  However, OpenAI APIs introduce unique challenges that differ from traditional APIs, such as the complexities of prompt engineering,  token-based cost management, non-deterministic outputs, and operation as black boxes. To the best of our knowledge, the challenges developers encounter when using OpenAI APIs have not been explored in previous empirical studies. To fill this gap, we conduct the first comprehensive empirical study by analyzing 2,874 OpenAI API-related discussions from the popular Q\&A forum Stack Overflow. We first examine the popularity and difficulty of these posts. After manually categorizing them into nine OpenAI API-related categories, we identify specific challenges associated with each category through topic modeling analysis. Based on our empirical findings, we finally propose actionable implications for developers, LLM vendors, and researchers.
\end{abstract}

}

\maketitle

\IEEEdisplaynontitleabstractindextext
\IEEEpeerreviewmaketitle

\section{Introduction}
\label{sec:introduction}

Large Language Models (LLMs) have significantly advanced the field of software engineering through their powerful natural language processing and generation capabilities, which allow them to understand, generate, and manipulate software artifacts (such as source code, code comments, bug reports, and Log files) effectively~\cite{fan2023large,hou2024large,wang2024software,zhang2023survey}. LLMs enable the automation of requirements analysis, code generation, software testing, and software maintenance, thereby improving the efficiency and quality of software development and maintenance processes.

To date, the most powerful LLMs are predominantly closed-source. These models are typically accessed via Application Programming Interfaces (APIs), allowing users to harness their advanced capabilities without the need for local deployment or extensive computational resources. Among the closed-source LLM providers, OpenAI stands out as one of the most prominent and influential LLM vendors, offering widely used models such as GPT-3.5 and GPT-4. These APIs have become integral to a variety of applications across domains, such as natural language processing, software development, education, healthcare, finance, and scientific research~\cite{sallam2023chatgpt}.

Compared to traditional programming language APIs, OpenAI APIs exhibit several important differences. Traditional APIs usually produce consistent and deterministic outputs for the same input, while OpenAI APIs may generate varying responses due to their probabilistic nature. Instead of relying on fixed parameters, developers must design prompts carefully to influence the behavior of OpenAI APIs. In terms of error handling, traditional APIs tend to provide explicit and structured error messages, whereas OpenAI APIs might return incorrect or misleading outputs without any indication of failure, requiring manual inspection and validation. Pricing models also differ significantly: traditional APIs often adopt clear and simple pricing schemes, while OpenAI APIs charge based on token consumption, which ties cost directly to the length of both inputs and outputs. Moreover, traditional APIs are typically well-documented and transparent, enabling developers to understand and predict their behavior. In contrast, OpenAI APIs operate more like black boxes, making it more difficult to anticipate their outputs or trace the source of unexpected results. These characteristics present unique challenges when using OpenAI APIs in software development. To the best of our knowledge, these challenges have not been thoroughly investigated in previous empirical studies.

To fill this gap, we conduct the first comprehensive empirical study to investigate the challenges faced by such APIs. In this study, we collect relevant discussions from Stack Overflow (SO), one of the most popular Q\&A forums. On SO, developers engage by posting questions and answers, allowing them to collaboratively tackle coding challenges, debug software faults, and share technical expertise~\cite{nugroho2024empirical,ahmad2018survey}.
To gather related posts, we first identified tags related to OpenAI APIs and selected initial posts based on these tags. We then applied manual inspection to further filter out irrelevant posts. As a result, we collected 2,874 posts.
Next, we categorized these posts into different API categories. In our empirical study, we mainly focus on nine primary categories based on previous suggestions~\cite{chen2021evaluating,roumeliotis2023chatgpt,chen2025empirical,auger2024overview}: Chat API, Image Generation API, Fine-tuning API, Embeddings API, Audio API, Code Generation API, Assistants API, GPT Actions API, and Others.

Based on the collected and organized data, we want to answer the following three research questions (RQs) in our empirical study.

\textbf{RQ1: What is the discussion trend on the OpenAI APIs among developers?}

\textbf{Results.} With the continuous development of OpenAI's large language models, discussions around the OpenAI API have shown a clear upward trend. However, due to shifts in developer behavior and some tensions within the community, there was a slight decline in such discussions in 2024.

\textbf{RQ2: How difficult is each category of OpenAI APIs?}

\textbf{Results.} 
To measure the difficulty of each category of OpenAI APIs, we considered two metrics. The first metric is the percentage of questions without accepted answers, and the second metric is the median time to receive an accepted answer. These metrics have been widely used in previous difficulty analyses of Stack Overflow posts
~\cite{mahmood2023empirical,tahaei2020understanding,process1,aly2021practitioners}.
The results indicate that questions related to the GPT Actions API are the most challenging, primarily because integrating GPT Actions requires developers to work with third-party APIs, which can be complex due to varying parameters and authentication methods. 
For other categories, general-purpose APIs (such as the Assistants API, Fine-tuning API, and Embeddings API) offer greater flexibility and broader functionality compared to specialized APIs (such as the Image API, Code Generation API, and Chat API). However, they also introduce higher complexity and greater challenges.

\textbf{RQ3: What are the challenges for each category of API?}

\textbf{Results.} 
Based on the analysis results, we observed that different categories of OpenAI APIs tend to reflect distinct topical focuses. For instance, the Chat API encompasses a wide range of nine topics, including model version migration, context management, and multimodal processing. In contrast, the Embeddings API is primarily associated with seven topics, such as vector database integration and retrieval-augmented generation.
When comparing these findings with challenges documented for traditional APIs~\cite{process1, venkatesh2016client, rosen2016mobile, alshangiti2019developing}, we identify several challenges that are unique to OpenAI APIs.
One prominent challenge concerns prompt design, particularly in the Chat API, Assistants API, and Code Generation API. Developers frequently seek guidance on how to apply prompt engineering techniques to improve the quality of generated conversations or code.
Another key issue relates to the cost of API usage. For example, users of the Chat API and Audio API often engage in discussions focused on optimizing token consumption
In addition, developers encounter task-specific challenges. Those working with the Audio API raise questions regarding audio format conversion; developers utilizing the Fine-tuning API frequently inquire about fine-tuning strategies, such as parameter-efficient fine-tuning (PEFT).
Emerging technologies such as retrieval-augmented generation (RAG) also introduce new challenges. As the Embeddings API plays a central role in RAG workflows, developers frequently engage in discussions related to its implementation, integration, and optimization.
In the context of continuously improving model capabilities, the deprecation of OpenAI APIs or models becomes inevitable, making compatibility issues difficult to avoid.
Finally, developers encounter significant challenges when integrating OpenAI APIs with third-party tools. These issues are particularly pronounced in scenarios involving the Chat API, Assistants API, and GPT Actions API. For instance, establishing connections to external data sources or invoking external functions often proves to be complex and error-prone.

Based on our empirical findings, we propose a set of actionable implications. Specifically, for developers, they should focus on prompt optimization, cost management, and context handling. For LLM vendors, they should provide higher-quality documentation, strengthen version management, improve context handling for multi-turn interactions, and offer better cost optimization strategies. For researchers, they should work on building comprehensive knowledge bases and developing code quality assurance tools, such as API search, API misuse detection, and deprecated API detection. These efforts will help address existing challenges and significantly enhance the overall experience of using OpenAI APIs.

To the best of our knowledge, the main contributions of our empirical study can be summarized as follows:

\begin{itemize}
    \item We conduct the first empirical study to investigate the challenges developers face when using OpenAI APIs.

    \item By manually categorizing these posts into nine OpenAI API categories and applying topic modeling techniques, we identify common challenges for each category.
    
    \item Based on the key findings, we provide actionable implications for LLM developers, LLM vendors, and researchers, aiming to help them improve API design, enhance documentation support, and increase compatibility.
    
    \item To support future research, we have made the collected posts and analysis scripts available on GitHub\footnote{\url{https://github.com/HDKHK/OpenAI-API}}, enabling others to reproduce our empirical study.
\end{itemize}

% The remainder of our empirical studies is structured as follows:
% Section~\ref{sec:background} introduces the background of OpenAI development and the categorization of APIs.
% Section~\ref{sec:methdology} describes the methodology used in our empirical study, including the methods for data collection, data labeling, popularity trend analysis, and challenge identification through topic modeling.
% Section~\ref{sec:RQ1} presents the results of the popularity analysis.
% Section~\ref{sec:RQ2} presents the results of the difficulty analysis.
% Section~\ref{sec:RQ3} summarizes the development challenges faced by each API category after performing topic modeling analysis.
% Section~\ref{sec:discussions} provides actionable implications for LLM stakeholders and analyzes potential threats to our study.
% Section~\ref{sec:related} summarizes related studies and emphasizes the novelty of our study.
% Section~\ref{sec:conclusion} concludes our empirical study.

\section{Background}
\label{sec:background}

As a leading LLM vendor, OpenAI continuously pushes the boundaries of technological capabilities through its flagship models, such as GPT-4 and the DALL·E series. OpenAI offers a comprehensive suite of APIs that enable developers to seamlessly integrate advanced LLM functionalities into a wide range of applications. Through the OpenAI API, developers gain access to powerful LLMs that can not only handle natural language understanding and generation tasks but also perform more complex operations, such as code generation, image recognition, and text-to-speech conversion.

Following the suggestions of previous studies~\cite{chen2021evaluating,roumeliotis2023chatgpt,chen2025empirical,auger2024overview}, we classify OpenAI APIs into nine primary categories in our empirical study. We then briefly introduce each API category and its main supported features.

\begin{itemize}
    \item \textbf{Chat API.} This API allows users to generate text or participate in chat conversations using LLMs, such as GPT-4o and GPT-4o mini. It is designed to support a wide range of conversational applications, from customer service bots to interactive characters.
    \item \textbf{Image Generation API.} The primary image generation LLM offered by OpenAI is DALL·E, which can create images from textual descriptions, edit existing images, and generate variations.
    \item \textbf{Fine-tuning API.} This API enables users to customize the behavior of GPT models based on specific datasets. It allows training models with user-gathered data to tailor responses more closely to the downstream task.
    \item \textbf{Embeddings API.} This API provides numerical representations of text (embeddings) that capture semantic meanings. These embeddings can be used for various purposes, including semantic search, content discovery, and more sophisticated natural language understanding tasks.
    \item \textbf{Audio API.} OpenAI offers the Whisper model for speech recognition, converting audio input into text. For text-to-speech, OpenAI provides a Text-to-Speech model that turns input text into natural-sounding spoken audio.
    \item \textbf{Code Generation API.} This includes models such as Codex, which are capable of understanding and generating code. These models assist in software development by generating code snippets, explaining code, and more.
    \item \textbf{Assistants API.} This API allows developers to build AI assistants within their applications. An assistant has instructions and can leverage models, tools, and knowledge to respond to user queries.
    \item \textbf{GPT Actions API.} This API allows ChatGPT users to interact with external applications using natural language, easily making RESTful API calls. It supports data retrieval and performing actions in other applications.
    \item \textbf{Others.} Edge cases that do not fall under the above eight categories but involve other API functionalities.
\end{itemize}

\section{Methodology}
\label{sec:methdology}

\begin{figure*}[htbp]
    \centering
  \includegraphics[width=0.92\textwidth]{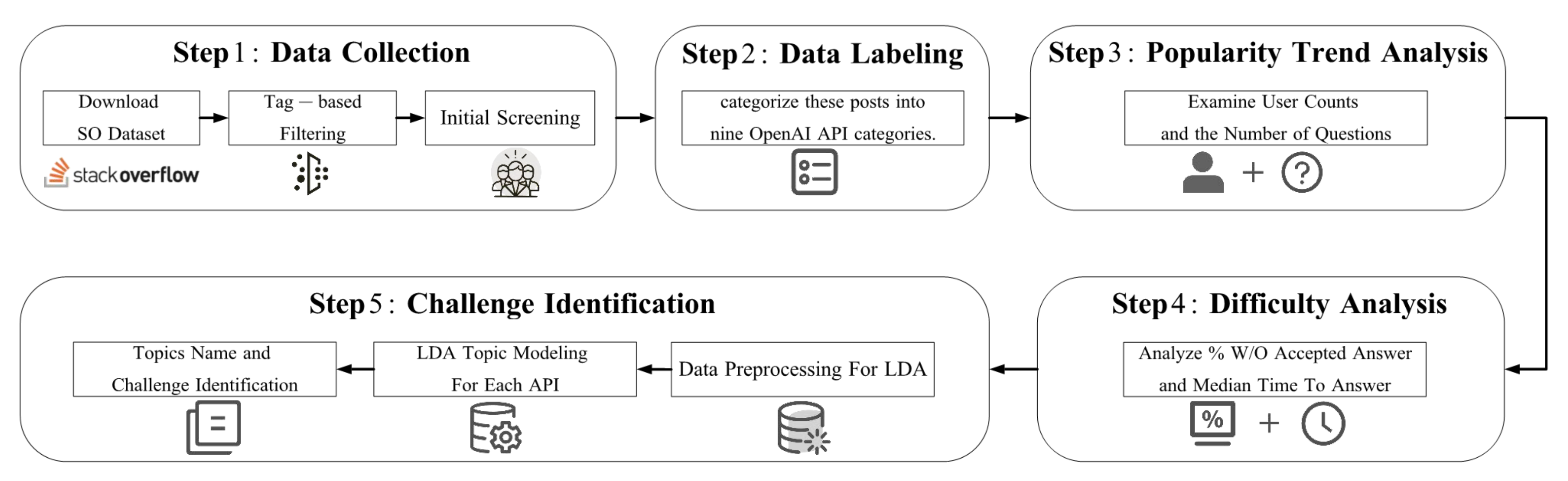} 
    \caption{Methodology overview of our empirical study.} % 图片标题
    \label{Fig:methodology} % 图片标签，用于引用
\end{figure*}

To identify the challenges developers face when using OpenAI APIs, we collect relevant questions from Stack Overflow. An overview of our research methodology is presented in Fig.~\ref{Fig:methodology}, which includes five main steps.
In the rest of this section, we show the details of these steps.

\subsection{Data Collection}

The rapid adoption of OpenAI APIs in software development has created a need to understand the popularity trends, difficulties, and challenges associated with their use. To address this, we systematically collect and curate a dataset of relevant SO posts to analyze developers' interactions with OpenAI APIs. 
Based on the data collection strategies in \cite{morovati2024common,shah2025towards,shah2025towards}, our data collection process involves three key steps: downloading the SO dataset, filtering posts based on relevant tags, and manually inspecting posts to ensure their relevance to OpenAI APIs.

\textbf{Step 1: Download $S_{so}$ Dataset.}
We downloaded the Stack Overflow dataset (denoted as $S_{so}$) from the Stack Exchange Data Dump\footnote{ \url{https://archive.org/details/stackexchange}, accessed on 22/1/2025}. This dataset contains posts up to January 22, 2025. For each post, we primarily extracted metadata including the user-submitted question, creation date, associated tags, title, body content, and the identifier of the accepted answer.

\textbf{Step 2: Tag-Based Filtering.}
Following previous studies~\cite{process2,significance-3,uddin2021empirical}, we identify a set of tags related to OpenAI APIs and extract the related posts from SO.
First, we define an initial set of tags, $T_0 = \langle \texttt{openai-api} \rangle$, to identify posts related to OpenAI APIs.
Second, we extract a post set $P$ from $S_{so}$, containing posts labeled with at least one tag from $T_0$. We define the OpenAI APIs tag set $T$ as the set of tags for the posts in $P$.
Third, by using the significance heuristic $\alpha$ and the relevance heuristic $\beta$, we measure the significance and relevance of each tag $t$ in $T$ and optimize $T$ by retaining tags that are significantly relevant to OpenAI APIs. The values of $\alpha$  and $\beta$  can be computed as follows.

\begin{equation}
    \text{Significance } \alpha = \frac{\text{Number of posts with tag } t \text{ in } P}{\text{Number of posts with tag } t \text{ in } S_{so}}
\end{equation}

\begin{equation}
    \text{Relevance } \beta = \frac{
        \text{Number of posts with tag } t \text{ in } P
    }{
        \text{Number of posts in } P
    }
\end{equation}

If the $\alpha$ and $\beta$ values of a tag $t$ are not lower than a pre-determined threshold, the tag is considered to have significant relevance~\cite{haque2020challenges}. By following the previous studies~\cite{significance-1,significance-2,significance-3},  we experiment with an extensive range of thresholds 
for $\alpha = \{0.05,\ 0.1,\ 0.15\,\ 0.2\,\ 0.25\,\ 0.3\,\ 0.35\}$ and $\beta = \{0.001,\ 0.005,\ 0.01\,\ 0.015\,\ 0.02\,\ 0.25\,\ 0.03\}$
and find that the most relevant set of OpenAI APIs tags can be identified when $\alpha$ is set to 0.1 and $\beta$ is set to 0.01. 
% The threshold we obtained is consistent with the values used in previous studies~\cite{significance-2,barua2014developers,significance-1,significance-3}.
Finally, we determined the following tag set $\langle \texttt{openai-api}, \texttt{chatgpt-api}, \texttt{langchain}, \texttt{fine-tuning}, \\ \texttt{openaiembeddings}  , \texttt{openai-whisper} \rangle$. 
We extract all posts that have at least one tag from $T$, resulting in 5,101 extracted posts.

\textbf{Step 3: Manual Inspection.}
Since users freely add tags to Stack Overflow, there may be cases of semantic ambiguity or cross-domain usage. For example, the tag $\langle \texttt{fine-tuning} \rangle$ can refer both to fine-tuning scenarios involving OpenAI APIs and to fine-tuning other models.
As a result, the posts retrieved through the Tag-Based Filtering step may include some that are unrelated to the development with OpenAI APIs.
Therefore, we need to conduct a manual inspection to identify posts of this type.
For example, the post ``How can I run a Python GitHub project in my Node.js project?"\footnote{\url{https://stackoverflow.com/questions/77792415}} focuses on inter-process communication between local Python and Node.js projects, rather than OpenAI APIs usage. Another example, ``How to fine-tune BERT Base (uncased model) for generating embeddings?"\footnote{\url{https://stackoverflow.com/questions/69943168}} pertains to local fine-tuning of BERT models for embedding generation, representing a self-training scenario using open-source pre-trained models (e.g., Hugging Face frameworks) without any reference to OpenAI APIs services.
In this step, three authors collaboratively conducted the process. The second and third authors independently inspected each post and removed those irrelevant to developers' use of OpenAI APIs.
We measured inter-rater reliability using Cohen’s Kappa ($\kappa$)\cite{cohen1960coefficient}, obtaining $\kappa = 0.832$, which indicates almost perfect agreement between annotators\cite{landis1977measurement}.
All discrepancies between the first two annotators were subsequently resolved through deliberation with the first author until full consensus was achieved.
As a result, a total of 2,874 valid posts that survived the manual inspection process proceeded to the data labeling step.

\subsection{Data Labeling}

\begin{figure}[htbp] % [htbp] 表示图片位置的浮动参数
    \centering % 居中显示图片
    \includegraphics[width=0.45\textwidth]{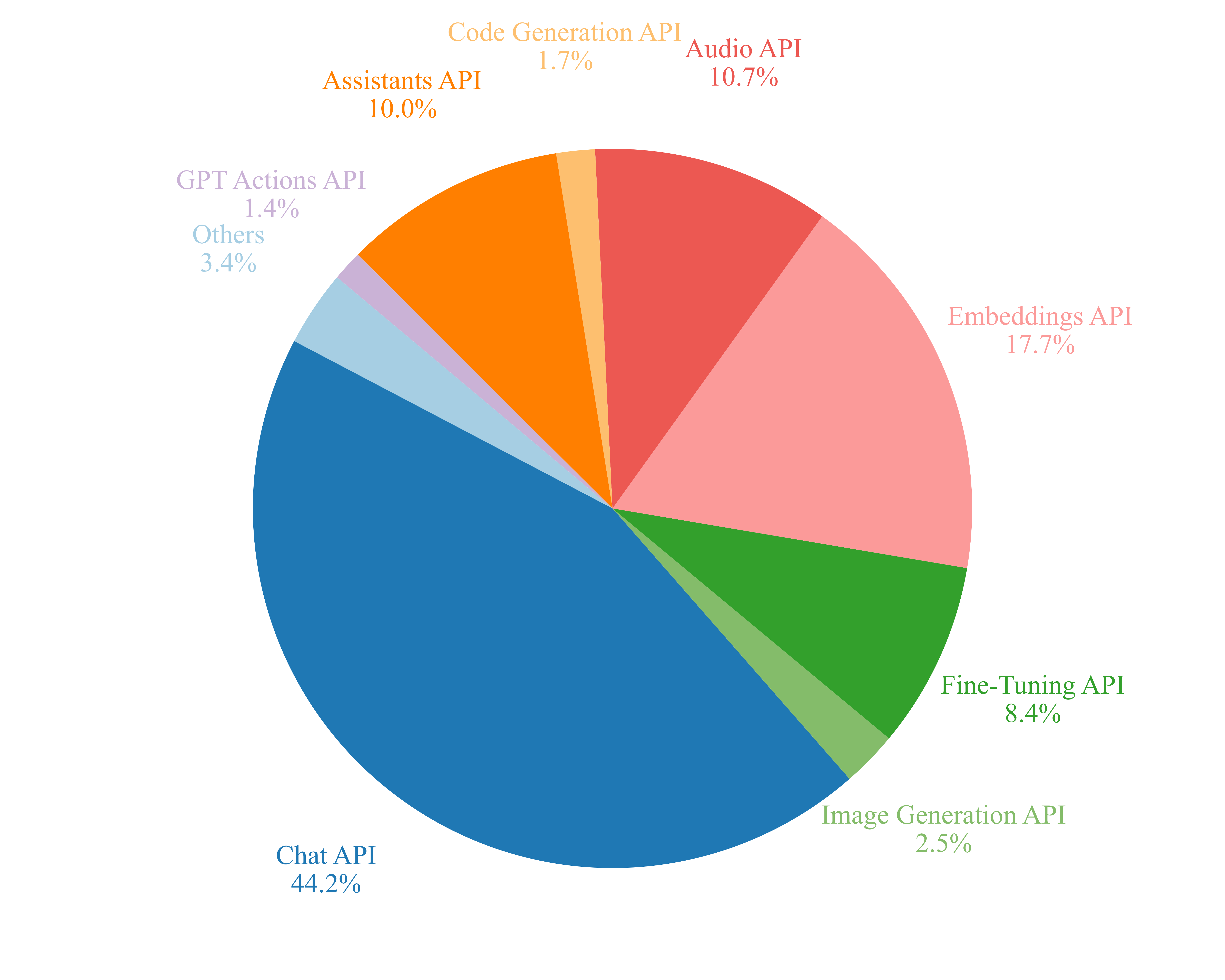} 
    \caption{Distribution of posts in nine OpenAI API categories.} % 图片标题
    \label{Fig:Distribution} % 图片标签，用于引用
\end{figure}

Based on the descriptions and definitions of OpenAI APIs, we label the posts and categorize them into nine OpenAI API categories.
For the data labeling step, we follow a methodology similar to that used in the manual inspection step. Specifically, the second and third authors independently classified each of the 2,874 valid posts, achieving an initial inter-rater reliability of Cohen’s Kappa ($\kappa$) = 0.822, indicating strong agreement between annotators.
All discrepancies were subsequently resolved through discussion with the first author until full consensus was reached.
The number of posts in each category is illustrated in Fig.~\ref{Fig:Distribution}.
Specifically, the Chat API dominates with 44.2\% of posts, underscoring its widespread adoption for conversational applications and likely reflecting its versatility and frequent use in diverse scenarios. The Embeddings API follows at 17.7\%, indicating substantial interest in semantic text representations for tasks like search and natural language understanding. The Audio API and Assistants API, with 10.7\% and 10.0\% respectively, 
also demonstrate a certain level of activity, suggesting growing developer focus on speech processing and AI assistant integration. Finally, the Fine-Tuning API (8.4\%), Others (3.4\%), Image Generation API (2.5\%), Code Generation API (1.7\%), and GPT Actions API (1.4\%) 
represent smaller proportions, possibly due to their more specialized use cases or limited demand.

\subsection{Popularity Trend Analysis}

We conduct a time series analysis to examine the popularity trends of OpenAI API usage among developers.
Specifically, we follow the methodology utilized in previous studies~\cite{alshangiti2019developing, shah2025towards, chen2025empirical} to compute the annual user counts and the number of questions related to this topic from 2021 to 2025.
The user counts represent the number of distinct users who actively engage with OpenAI API-related discussions by posting questions or answering them.
Notice that since some posts may contain extensive discussions, a single post may involve multiple users.
In summary, the annual user counts reflect changes in the user base. Meanwhile, the annual number of questions, which tracks the number of new posts each year, indicates the level of developer engagement and interest in this topic.
Therefore, these metrics can provide a comprehensive view of the popularity trend.

\subsection{Difficulty Analysis}
By evaluating the difficulty levels of questions across the nine OpenAI API categories, we can pinpoint the specific APIs where developers face significant challenges. 
Moreover, identifying the more challenging categories can help prioritize efforts to enhance documentation support and improve the APIs for these categories~\cite{haque2020challenges, process2}.
To achieve this goal, we apply two widely used metrics from prior studies~\cite{significance-3,alshangiti2019developing,significance-1} to measure the difficulty level of posts in nine major OpenAI API categories: the percentage of questions without accepted answers (\% w/o accepted answers) and the median time to receive an accepted answer (Median Time to Answer). Notice that the time to receive an accepted answer is the creation time of the answer, not when the answer is marked as accepted.  By combining these two metrics, we can obtain a more comprehensive view of the difficulty level of OpenAI API-related questions.

\subsection{Challenge Identification}

Identifying the challenges developers encounter when using OpenAI APIs is crucial for improving usability, refining documentation, and strengthening developer support systems. To this end, we apply topic modeling to identify the topics associated with each OpenAI API category, enabling a deeper understanding of the challenges developers face. This topic identification approach has been widely adopted in previous studies~\cite{mahmood2023empirical, tahaei2020understanding, process1, aly2021practitioners} and has been shown to effectively extract and categorize topics. Our challenge identification includes the following three steps: data preprocessing, topic identification, and topic name identification.

\textbf{Step 1: Data Preprocessing.}
In this step, we first extract only the post titles and exclude the body content for two reasons: body text may introduce analytical noise~\cite{process1,process2}, and post titles have been empirically validated as representative of body content~\cite{chen2016learning, xu2017answerbot}. This setting has also been adopted in previous studies~\cite{process1,process2, rosen2016mobile}. Next, we remove stopwords (e.g., ‘you,’ ‘we,’ and ‘our’) using the NLTK stopwords corpus. Finally, we apply stemming and lemmatization to standardize the extracted tokens to their root forms (e.g., ‘running’ becomes ‘run’).

% \wjb{topic identification}
\textbf{Step 2: Topic Identification.}
In our study, we consider $K$ values ranging from 2 to 20 (in increments of 1) and calculate their corresponding coherence scores, which assess the degree of semantic similarity among the top words in each topic and are widely used as a measure of topic interpretability and quality~\cite{roder2015exploring}. After selecting the $K$ value with the optimal coherence score, we validate the result by randomly sampling 30 posts per topic within the range [max($K$–8, 1), $K$+8]. When fewer than 30 posts are available for a topic, all available posts are included. We ultimately set the range size to 8 based on considerations of stability and computational efficiency. Specifically, coherence scores within this range show only minor variation, and a range of 8 is sufficiently wide to capture variations in topic quality around the optimal $K$.
During validation, two authors manually inspect the 30 sampled posts for each $K$ value in the range. Following previous studies~\cite{process2,lda1}, we apply the following evaluation criteria: (1) \textbf{clarity.} the posts align with a clear and identifiable topic; (2) \textbf{specificity.} the topic is sufficiently distinct to exclude unrelated content; and (3) \textbf{generalizability.} the topic represents a meaningful category of related posts.

\textbf{Step 3: Topic Name and Challenge Identification.}
We follow the methodology used in previous studies~\cite{significance-3, chen2020comprehensive, lda1} and apply the open card sort method~\cite{fincher2005making} to determine the topic names. For each topic, we manually examine the top 20 words in its word set~\cite{lda1, process2} to select a topic name that best represents the topic and its associated posts. These top 20 words are the most probable and representative terms for the topic. 
Additionally, we randomly sample and review 30 posts related to the topic~\cite{uddin2021empirical, process2} to verify whether the chosen topic name accurately reflects the posts, and refine the name if necessary. During this process, we also use the open card sort method~\cite{fincher2005making} to manually merge semantically similar topics. Specifically, the first three authors independently assign names to the topics and then engage in iterative discussions to refine the topic names until a consensus is reached.

\section{RQ1: Popularity Analysis}
\label{sec:RQ1}

\begin{figure}[htbp] % [htbp] 表示图片位置的浮动参数
    \centering % 居中显示图片
    \includegraphics[width=0.45\textwidth]{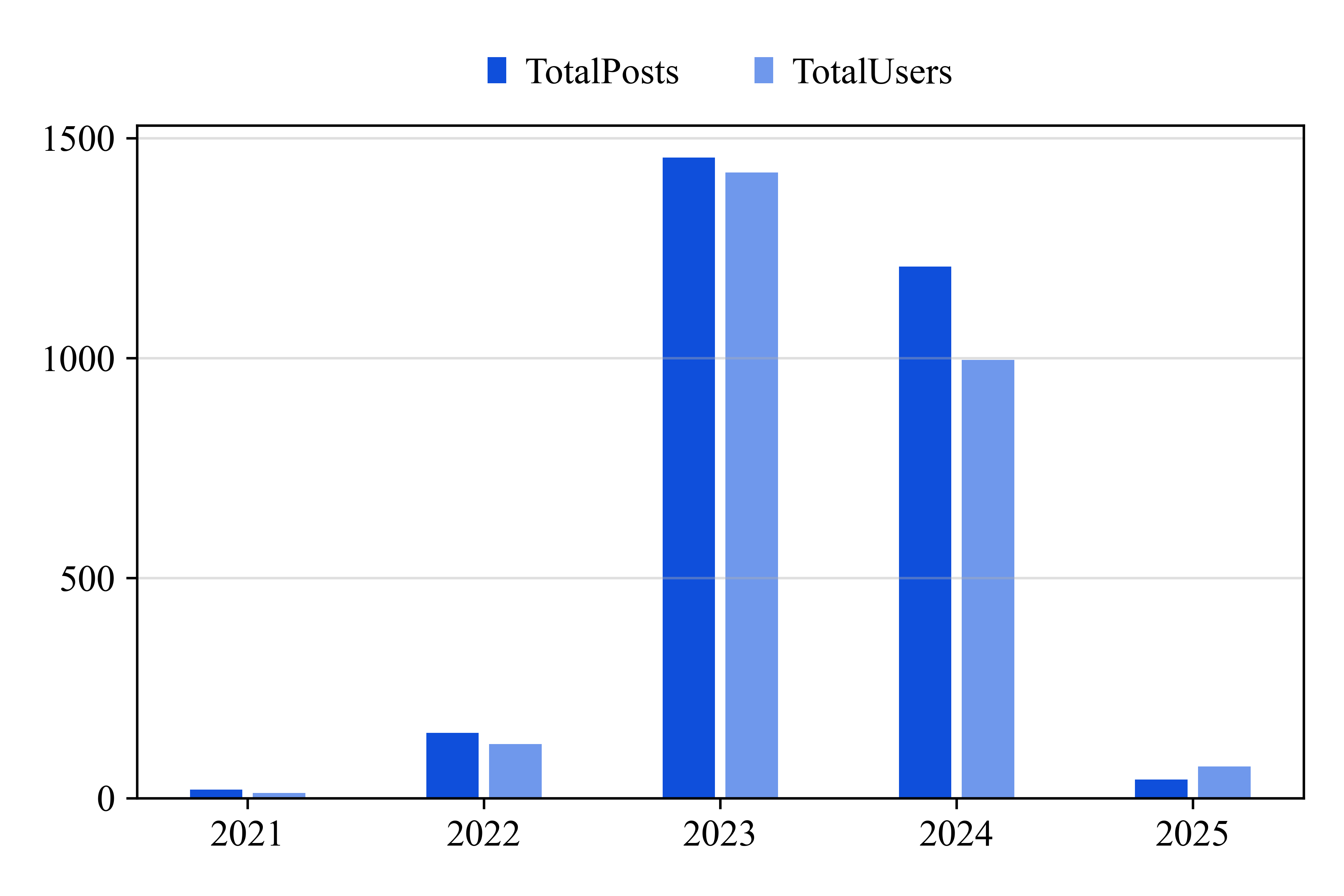} 
    \caption{The number of posts and users related to OpenAI APIs over time.} % 图片标题
    \label{Fig:number} % 图片标签，用于引用
\end{figure}

Fig.~\ref{Fig:number} shows the popularity trend for OpenAI APIs discussions in terms of the number of users and questions on Stack Overflow.
In this figure, we find that from 2021 to January 22, 2025, discussions related to OpenAI APIs on Stack Overflow exhibited an overall upward trend, with notable growth in both the number of posts and participating users.

Specifically, from 2021 to 2022, discussions on Stack Overflow related to OpenAI APIs remained limited. In 2021, there were only 19 posts and 12 users participating in these topics. Although 2022 saw a modest increase, with 148 posts and 123 users, overall engagement was still relatively low. This limited activity can be attributed to the early phase of OpenAI APIs. Officially released in late 2020, the API was still in its initial adoption period during 2021 and 2022. Developers were primarily exploring its capabilities privately or within smaller communities, rather than engaging in public discussions on the popular forum like Stack Overflow.

In contrast, 2023 witnessed a sharp rise in discussions about OpenAI. This was evident in the significant growth in both the number of posts and users, which reached 1,456 posts and 1,422 users. The increase was largely driven by the rapid adoption of generative AI tools such as ChatGPT. With growing interest in AI-assisted development, more developers began asking questions, sharing experiences, and participating in technical discussions related to OpenAI’s APIs and tools. This shift reflected the broader incorporation of AI technologies into common software development practices.

Though 2023 experienced particularly high activity, discussions continued at a notable pace in 2024, with 1,208 posts and 996 users. This still represents a substantial volume of activity compared to previous years, despite a slight decrease. This variation can be attributed to two main factors. On one hand, the widespread use of generative AI tools like ChatGPT and GitHub Copilot allowed developers to obtain quick code suggestions and solutions directly from these tools, reducing their reliance on traditional platforms such as Stack Overflow~\cite{del2024large}. On the other hand, Stack Overflow’s partnership with OpenAI triggered concern among some community members. Several users were worried that their contributions were being used to train AI models without proper consent, which led some of them to delete or modify their previously posted content.

In 2025, although the data is only available up to January 22, 2025, there have been 42 posts and 72 users during this period, indicating that the engagement with OpenAI API-related discussions on Stack Overflow is relatively stable.
% 删除了 and still on an upward trend

% 设置颜色、背景、边框等
\begin{tcolorbox}[
    colback=gray!10,    % 背景色
    colframe=black,   % 边框颜色
    boxrule=0.6pt,  
    sharp corners       % 直角边框
]
\textbf{Finding 1.} As the OpenAI API matured and its real-world applications increased, technical discussions on Stack Overflow initially grew, indicating a broader acceptance of LLMs in software development and attracting more developers to engage in discussions about the OpenAI API. However, developer behavior shifts and community tensions have led to a recent decline in participation.
\end{tcolorbox}

\section{RQ2: Difficulty Analysis}
\label{sec:RQ2}

%参考What Do Concurrency Developers Ask About_A Large-Scale Study on Stack Overflow
Table~\ref{Table:difficulty} presents the percentage of questions without accepted answers (second column) and the median time to receive an accepted answer, measured in hours (third column), for each OpenAI API category. The table is sorted in descending order based on the percentage of questions without accepted answers.
% Additionally, it provides the total number of related posts and the number of posts without accepted answers for each category.
% In our empirical study, we consider an API category to be more difficult if it has a higher percentage of unanswered questions or a longer median time to receive an accepted answer.

% results...
\begin{table}[htbp]
    \centering
    \caption{\textbf{The difficulty with each OpenAI API category.}}
    \label{Table:difficulty}
    \begin{tabular}{lcc}
    \toprule%第一道横线
        \multirow{2}{*}{\textbf{OpenAI API Category}} & \textbf{Posts w/o}
 & \textbf{Median}\\ 
        &\textbf{Accepted (\%)}&\textbf{Time (h)}\\ 
        \midrule%第二道横线 
        GPT Actions API & 94.9 & 583.3 \\ 
        Assistants API & 83.0 & 39.9 \\ 
        Audio API & 80.1 & 62.0 \\ 
        Embeddings API & 79.4 & 15.8 \\ 
        Fine-Tuning API & 79.3 & 36.3 \\ 

        Others & 78.6 & 28.3 \\ 
        Chat API & 74.9 & 5.6 \\ 
          Code Generation API & 72.0 & 3.9 \\
        Image API & 68.1 & 8.4 \\ 
       \bottomrule%第四道横线
    \end{tabular}
\end{table}

According to Table~\ref{Table:difficulty}, the GPT Actions API is considered the most challenging due to several factors. First, integrating GPT Actions often requires developers to interact with third-party APIs, which can be complex due to varying parameters and authentication methods. This complexity increases the likelihood of errors and necessitates a deeper understanding of both the GPT Actions framework and the external APIs involved. Second, developers have reported issues such as GPT Actions making multiple redundant API calls, ignoring instructions, and experiencing slow response times. These issues are particularly challenging because they not only complicate debugging and maintenance but also make it difficult to identify the root cause, often requiring extensive investigation and testing to resolve.

General-purpose APIs, such as the Assistants API, Fine-tuning API, and Embeddings API, often pose more challenges than specialized APIs like the Image API, Code Generation API, and Chat API. This is because general-purpose APIs are designed for more complex tasks and must handle a wide variety of inputs and outputs. 
For example, the Assistants API enables the development of multifunctional AI assistants that interact with multiple external systems and handle diverse data types, including text, voice, and files, thereby significantly increasing the complexity of both development and debugging.
Additionally, general-purpose APIs are applied across a broad range of use cases, often involving diverse business logic and user requirements, which leads to more edge cases and exceptions that require customization and fine-tuning. In contrast, specialized APIs are optimized for specific tasks, such as image generation, code generation, or chat interactions, making them more focused, easier to understand, and generally less prone to issues.

% 设置颜色、背景、边框等
\begin{tcolorbox}[
    colback=gray!10,    % 背景色
    colframe=black,   % 边框颜色
    boxrule=0.6pt,  
    sharp corners       % 直角边框
]
\textbf{Finding 2.} The GPT Actions API is considered the most challenging due to its complexity in integrating with third-party APIs and unpredictable behaviors. For other categories, compared to specialized APIs, general-purpose APIs offer advantages in flexibility and broad functionality, but they also come with increased complexity and challenges.
\end{tcolorbox}

\section{RQ3: Challenge Identification}
\label{sec:RQ3}

Table~\ref{Fig:topic} presents the topic names, keywords, and categorization corresponding to each OpenAI API category. 
These keywords are selected from the top 20 high-frequency words, based on their representativeness and determined through manual review and discussion. Finally, we manually analyze each post to summarize the specific challenges associated with each category~\cite{alam2024developer, process1, haque2020challenges}.
Based on this table, we observe that the topics covered by different Open API categories vary to some extent. In the remainder of this section, we introduce the topics associated with each category, using representative posts to illustrate the topics they cover.

\begin{table*}[!b]
    \centering
    \caption{Topic names and corresponding keywords for nine OpenAI API categories.}
    \label{Fig:topic} % 图片标签，用于引用
    \begin{tabular}{
  >{\RaggedRight\arraybackslash}m{75pt}
  >{\RaggedRight\arraybackslash}m{100pt}
  >{\centering\arraybackslash}m{330pt}
}
\toprule
\textbf{OpenAI API Category} & \textbf{Topic Name} & \textbf{Keywords} \\
\midrule
\multirow{16}{*}{Chat API}  
  & A1: API Core Operation Errors &
  \texttt{typeerror}, \texttt{function}, \texttt{sdk}, \texttt{chatcomplet}, \texttt{react}, \texttt{js}, \texttt{stream}, \texttt{unabl} \\
  & A2: Model Migration & 
  \texttt{token}, \texttt{exceed}, \texttt{llm}, \texttt{modul}, \texttt{attributeerror}, \texttt{quota}, \texttt{valid}, \texttt{paramet} \\
  & A3: Context Management & 
  \texttt{model}, \texttt{prompt}, \texttt{code}, \texttt{langchain}, \texttt{question}, \texttt{process}, \texttt{davinci}, \texttt{support} \\
  & A4: Streaming and Asynchronous Processing & 
  \texttt{stream}, \texttt{request}, \texttt{respons}, \texttt{javascript}, \texttt{timeout}, \texttt{fastapi}, \texttt{async}, \texttt{convers} \\
   &A5: Security and Authorization & 
  \texttt{endpoint}, \texttt{deploy}, \texttt{document}, \texttt{variabl}, \texttt{context}, \texttt{messag}, \texttt{custom}, \texttt{access} \\         
   &A6: Performance Optimization & 
  \texttt{respons}, \texttt{node}, \texttt{memory}, \texttt{input}, \texttt{fix}, \texttt{json}, \texttt{langchain}, \texttt{pars} \\
   &A7: Multimodal Processing & 
  \texttt{file}, \texttt{data}, \texttt{extract}, \texttt{web}, \texttt{html}, \texttt{load}, \texttt{applic}, \texttt{resourc} \\  
   &A8: Framework and Toolchain Integration & 
  \texttt{key}, \texttt{invalid}, \texttt{generat}, \texttt{post}, \texttt{url}, \texttt{turbo}, \texttt{complet}, \texttt{open} \\        
   &A9: Custom Functionality & 
  \texttt{implement}, \texttt{app}, \texttt{content}, \texttt{return}, \texttt{connect}, \texttt{problem}, \texttt{async}, \texttt{local} \\    
  \midrule
\multicolumn{3}{r}{\textit{Continued on next page}} \\
    \end{tabular} 
  
\end{table*}

\begin{table*}[!t]
    \centering 
    \begin{tabular}{
  >{\RaggedRight\arraybackslash}m{75pt}
  >{\RaggedRight\arraybackslash}m{100pt}
  >{\centering\arraybackslash}m{330pt}
}
\multicolumn{3}{c}{\bf\normalfont\sffamily TABLE~\ref{Fig:topic}}\\
\multicolumn{3}{c}{\bf\normalfont\sffamily Topic names and corresponding keywords for nine OpenAI API categories (continued)}\\ [3ex]
\toprule
\textbf{OpenAI API Category} & \textbf{Topic Name} & \textbf{Keywords} \\                   
\midrule
\multirow{7}{*}{Embeddings API} 
  &B1: Vector Database Integration and Maintenance & 
  \texttt{langchain}, \texttt{document}, \texttt{data}, \texttt{faiss}, \texttt{chromadb}, \texttt{db}, \texttt{vector}, \texttt{memori} \\
  &B2: API Request Failures and Rate Limitation & 
  \texttt{pinecon}, \texttt{file}, \texttt{respons}, \texttt{token}, \texttt{openaiembed}, \texttt{vectorstor}, \texttt{csv}, \texttt{limit} \\              
  &B3: Attribute and Query Errors & 
  \texttt{queri}, \texttt{attribut}, \texttt{return}, \texttt{similar}, \texttt{emb}, \texttt{generat}, \texttt{score}, \texttt{calcul} \\
  &B4: Retrieval-Augmented Generation & 
  \texttt{rag}, \texttt{index}, \texttt{llm}, \texttt{prompt}, \texttt{vector}, \texttt{chunk}, \texttt{metadata}, \texttt{llamaindex} \\
  &B5: Embedding Model Configuration & 
  \texttt{embed}, \texttt{search}, \texttt{azur}, \texttt{gpt}, \texttt{semant}, \texttt{dimens}, \texttt{context}, \texttt{cousin} \\       
  &B6: Data Management and Storage & 
  \texttt{text}, \texttt{chroma}, \texttt{model}, \texttt{pdf}, \texttt{load}, \texttt{chromadb}, \texttt{exist}, \texttt{issue} \\
  &B7: Advanced Applications & 
  \texttt{vector}, \texttt{store}, \texttt{creat}, \texttt{python}, \texttt{function}, \texttt{modul}, \texttt{import}, \texttt{chat} \\
\midrule
\multirow{6}{*}{Audio API} 
  &C1: Model Errors and File Loading & 
  \texttt{whisper}, \texttt{audio}, \texttt{file}, \texttt{transcrib}, \texttt{attributeerror}, \texttt{numpi}, \texttt{requir}, \texttt{system} \\
  &C2: Format Conversion and Stream Processing & 
  \texttt{format}, \texttt{ffmpeg}, \texttt{convert}, \texttt{wav}, \texttt{stream}, \texttt{respons}, \texttt{modul}, \texttt{call} \\
  &C3: Cross-Platform Deployment & 
  \texttt{instal}, \texttt{react}, \texttt{nativ}, \texttt{gpu}, \texttt{app}, \texttt{languag}, \texttt{code}, \texttt{transcript} \\
  &C4: API Request Parameter Errors & 
  \texttt{request}, \texttt{timestamp}, \texttt{js}, \texttt{nodej}, \texttt{type}, \texttt{object}, \texttt{load}, \texttt{integr} \\
  &C5: Performance Optimization and Cost Management & 
  \texttt{token}, \texttt{cuda}, \texttt{directori}, \texttt{process}, \texttt{version}, \texttt{asr}, \texttt{invalid}, \texttt{generate} \\
  &C6: Large-scale File Processing & 
  \texttt{audio}, \texttt{transcrib}, \texttt{chunk}, \texttt{video}, \texttt{filenotfounderror}, \texttt{import}, \texttt{run}, \texttt{client} \\       
\midrule
\multirow{6}{*}{Assistants API} 
  &D1: Core Functions and Fundamental Configuration &
  \texttt{agent}, \texttt{model}, \texttt{function}, \texttt{code}, \texttt{format}, \texttt{document}, \texttt{file}, \texttt{assist} \\
  &D2: Context Maintenance and Enhancement &
  \texttt{convers}, \texttt{custom}, \texttt{chat}, \texttt{question}, \texttt{input}, \texttt{respons}, \texttt{user}, \texttt{system} \\
  &D3: Tool Integration and Model Extension &
  \texttt{llm}, \texttt{tool}, \texttt{integr}, \texttt{json}, \texttt{upload}, \texttt{rag}, \texttt{langchain}, \texttt{output} \\
  &D4: Real-time Response and Stream Processing &
  \texttt{stream}, \texttt{respons}, \texttt{websocket}, \texttt{handl}, \texttt{run}, \texttt{react}, \texttt{chatbot}, \texttt{assist} \\
  &D5: Threads and Automation Processing &
  \texttt{thread}, \texttt{messag}, \texttt{process}, \texttt{run}, \texttt{chain}, \texttt{retriev}, \texttt{data}, \texttt{file} \\
  &D6: Function Calling &
  \texttt{function}, \texttt{call}, \texttt{paramet}, \texttt{argument}, \texttt{schema}, \texttt{type}, \texttt{python}, \texttt{prompt} \\
\midrule
\multirow{5}{*}{Fine-tuning API} 
  &E1: Basic Error Handling &
  \texttt{error}, \texttt{expect}, \texttt{found}, \texttt{face}, \texttt{batch}, \texttt{size}, \texttt{generat}, \texttt{openai} \\
  &E2: Customization via Model Fine-tuning &
  \texttt{llama}, \texttt{data}, \texttt{transform}, \texttt{prompt}, \texttt{differ}, \texttt{finetun}, \texttt{complet}, \texttt{result} \\
  &E3: Dataset Construction and Output Control &
  \texttt{custom}, \texttt{dataset}, \texttt{input}, \texttt{classif}, \texttt{task}, \texttt{peft}, \texttt{answer}, \texttt{return} \\
  &E4: API Basic Usage &
  \texttt{use}, \texttt{file}, \texttt{format}, \texttt{gpu}, \texttt{creat}, \texttt{turbo}, \texttt{type}, \texttt{api} \\
  &E5: Model Management and Resource Optimization &
  \texttt{python}, \texttt{upload}, \texttt{token}, \texttt{code}, \texttt{loss}, \texttt{output}, \texttt{base}, \texttt{api} \\
\midrule
\multirow{3}{*}{Image Generation API} 
  &F1: Image Input Formats and API Configuration &
  \texttt{imag}, \texttt{api}, \texttt{rgba}, \texttt{upload}, \texttt{configur}, \texttt{request}, \texttt{js}, \texttt{error} \\
  &F2: API Usage Limits and Cross-Environment Invocation Conflicts &
  \texttt{generat}, \texttt{error}, \texttt{limit}, \texttt{firebas}, \texttt{call}, \texttt{fail}, \texttt{variat}, \texttt{use} \\
  &F3: Generated Image Processing and Version Updates &
  \texttt{dall}, \texttt{url}, \texttt{download}, \texttt{nodej}, \texttt{python}, \texttt{creat}, \texttt{librari}, \texttt{implement} \\
\midrule
\multirow{3}{*}{Code Generation API} 
  &G1: Fundamental API Usage and Integration Issues &
  \texttt{api}, \texttt{openai}, \texttt{python}, \texttt{sdk}, \texttt{async}, \texttt{return}, \texttt{function}, \texttt{code} \\
  &G2: Code Generation &
  \texttt{code}, \texttt{error}, \texttt{sql}, \texttt{langchain}, \texttt{prompt}, \texttt{complet}, \texttt{php}, \texttt{suffix} \\
  &G3: Model Parameters and Output Control &
  \texttt{codex}, \texttt{model}, \texttt{token}, \texttt{output}, \texttt{argument}, \texttt{prefix}, \texttt{text}, \texttt{complet} \\
\midrule
\multirow{2}{*}{GPT Actions API} 
  &H1: Custom Action and Multi-Operation Management &
  \texttt{action}, \texttt{custom}, \texttt{function}, \texttt{agent}, \texttt{integr}, \texttt{langchain}, \texttt{tool}, \texttt{llm} \\
  &H2: External Data Connectivity and File Processing &
  \texttt{data}, \texttt{file}, \texttt{access}, \texttt{schema}, \texttt{server}, \texttt{googl}, \texttt{send}, \texttt{configur} \\
\midrule
\multirow{2}{*}{Others} 
  &I1: General Technical Barriers &
  \texttt{error}, \texttt{rate}, \texttt{key}, \texttt{azur}, \texttt{token}, \texttt{api}, \texttt{openai}, \texttt{proxi} \\
  &I2: OpenAI API Edge Cases &
  \texttt{content}, \texttt{request}, \texttt{model}, \texttt{url}, \texttt{post}, \texttt{limit}, \texttt{ssl}, \texttt{file}  \\
  \bottomrule
    \end{tabular} 
   
\end{table*}

\subsection{Identified Topics for Chat API}

The Chat API, a core component of OpenAI's API ecosystem, empowers developers to embed conversational AI capabilities into their applications. In our classification, this category accounts for 44.2\% of all developer discussions. Through topic analysis, we identify nine distinct topics, which are introduced as follows.

\begin{figure}[htbp] % [htbp] 
\centering   
\includegraphics[width=0.45\textwidth]{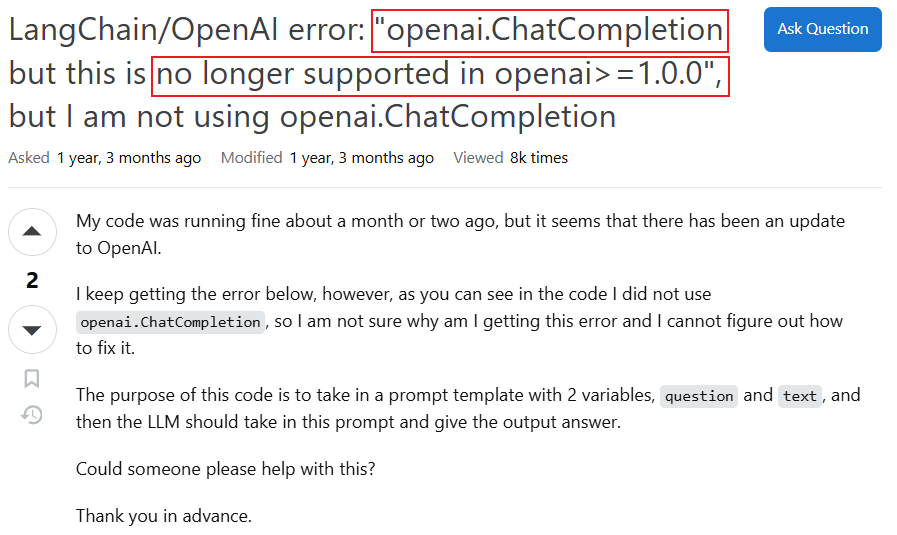}    
\caption{A sample post in the topic of API core operation errors.}   
\label{Fig:API_Core_Functionality_Errors_Sample}
\end{figure}

\textbf{A1: API Core Operation Errors.} 
This topic explores errors in OpenAI's Chat API that hinder core operations, focusing on three primary issues:
First, updates to Software Development Kits (SDKs) may cause function deprecation. For example, with the release of OpenAI Python SDK version 1.0.0, the \texttt{openai.ChatCompletion} method was deprecated. Developers who had not updated their codebases encountered compatibility issues\footnote{\url{https://stackoverflow.com/questions/77540822}} as shown in Fig.~\ref{Fig:API_Core_Functionality_Errors_Sample}.
Second, there may be anomalies in the streaming protocol. For instance, users reported inconsistencies when utilizing the Chat API’s streaming capabilities. These issues include duplicated outputs and unexpected interruptions in the data stream\footnote{\url{https://stackoverflow.com/questions/76125712}}.
Third, API Key Authentication may cause failures in integrated systems. For example, in environments like \texttt{CrewAI}, developers face challenges where valid OpenAI API keys were erroneously rejected. This problem particularly arises when integrating with alternative models or platforms, such as Hugging Face or Ollama\footnote{\url{https://stackoverflow.com/questions/78685685}}.

\begin{figure}[htbp] % [htbp] 表示图片位置的浮动参数    
\centering % 居中显示图片    
\includegraphics[width=0.45\textwidth]{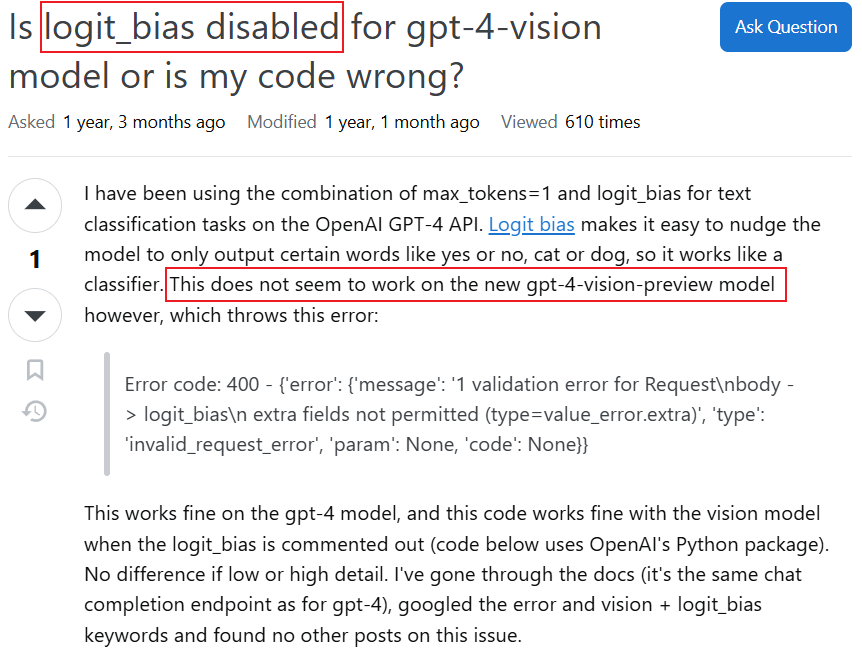}    
\caption{A sample post in the topic of model migration.} % 图片标题    
\label{Fig:Model_Version_Migration_Sample} % 图片标签，用于引用
\end{figure}

\textbf{A2: Model Migration.} 
This topic focuses on the challenges encountered during the migration of OpenAI models and SDK versions. One of the main issues is the deprecation of legacy models, which forces developers to adapt their workflows to newer models. For example, migrating from \texttt{text-davinci-003} to \texttt{gpt-3.5-turbo} presents challenges in code modification\footnote{\url{https://stackoverflow.com/questions/75774552}}. 
In addition, when upgrading the OpenAI Node.js SDK from version 3 to version 4, developers may experience invocation failures due to changes in API initialization methods and model deprecations\footnote{\url{https://stackoverflow.com/questions/77807093}}. 
Parameter \texttt{schemas}' inconsistency further complicates the migration process, especially when using specialized models. As illustrated in Fig.~\ref{Fig:Model_Version_Migration_Sample}, the GPT-4 vision model lacks support for \texttt{logit\_bias}, resulting in unexpected behavioral deviations\footnote{\url{https://stackoverflow.com/questions/77564810}}.

% Results
\textbf{A3: Context Management.}
This topic centers on the challenge of balancing multi-turn dialogue dynamics with the inherent context length limitations of large language models. A primary concern is the management of dialogue history, where developers seek effective strategies to maintain contextual coherence across multiple interactions (e.g., managing context in \texttt{gpt-3.5-Turbo}\footnote{\url{https://stackoverflow.com/questions/75710916}}).
The second key issue involves token limitations, especially when the length of input and output sequences exceeds the model’s maximum context length. In such cases, optimizing prompts to avoid truncation and ensure smooth dialogue becomes particularly important\footnote{\url{https://stackoverflow.com/questions/70060847}}.
Finally, handling lengthy inputs poses a significant challenge. Developers often preprocess or segment large or structurally complex texts to meet API constraints and preserve information integrity\footnote{\url{https://stackoverflow.com/questions/75777566}}.

\begin{figure}[htbp] % [htbp] 
\centering  
\includegraphics[width=0.45\textwidth]{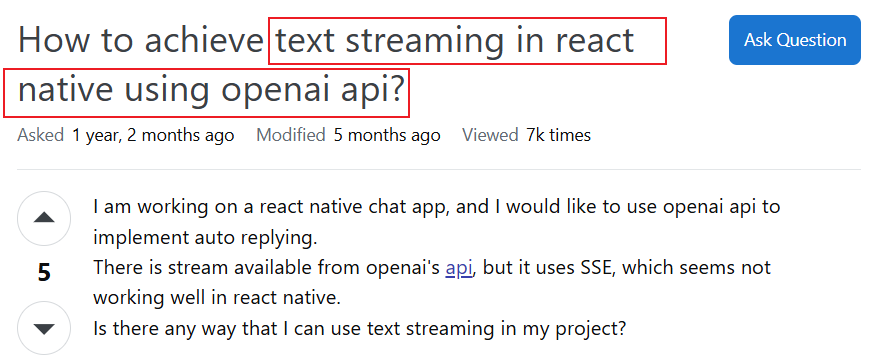}    
\caption{A sample post in the topic of streaming and asynchronous processing.}   
\label{Fig:Streaming_and_Asynchronous_Processing_Sample} 
\end{figure}

\textbf{A4: Streaming and Asynchronous Processing.}
This discussion centers on the technical implementation of real-time response handling and asynchronous operations in OpenAI's Chat API integrations. A core challenge involves effectively incorporating streaming APIs with frontend frameworks like \texttt{React} to facilitate seamless text streaming in applications.
As illustrated in Fig.~\ref{Fig:Streaming_and_Asynchronous_Processing_Sample}, developers frequently explore approaches to implement real-time text streaming in \texttt{React} Native environments using the Chat API\footnote{\url{https://stackoverflow.com/questions/77725698}}. Another critical consideration involves optimizing asynchronous API calls through performance-enhancing tools such as Python's \texttt{asyncio} and \texttt{aiohttp} libraries\footnote{\url{https://stackoverflow.com/questions/75805772}}, which enable efficient concurrent request processing.
Finally, practical implementation challenges arise in data management, particularly when persisting chunked streaming responses into structured storage systems such as chat history databases\footnote{\url{https://stackoverflow.com/questions/79149341}}.

\textbf{A5: Security and Authorization.}
This topic focuses on three primary security and permission challenges commonly encountered in Chat API implementations.
The first challenge involves the risk of API key exposure, often caused by misconfigured environment variables that inadvertently make keys publicly accessible\footnote{\url{https://stackoverflow.com/questions/77797590}}.
The second challenge pertains to Cross-Origin Resource Sharing (CORS) restrictions, where frontend applications are unable to complete API requests due to strict CORS policies\footnote{\url{https://stackoverflow.com/questions/77639308}}.
The third issue concerns insufficient permissions, exemplified by 401 authentication errors that arise when API calls are made without proper authorization\footnote{\url{https://stackoverflow.com/questions/75827468}}.

\textbf{A6: Performance Optimization.}
This topic focuses on performance optimization challenges in Chat API implementations, specifically addressing resource constraints and enhancing response efficiency.
The first issue involves rate limits, where developers face restrictions on the frequency of API requests. This is exemplified by error cases such as rate limit errors encountered when using the Azure chat playground with data sources and system messages\footnote{\url{https://stackoverflow.com/questions/78899976}}.
The second challenge concerns optimizing token usage, where developers aim to reduce costs by minimizing input and output lengths. An example is the query on how to reduce the token usage when sending data to ChatGPT\footnote{\url{https://stackoverflow.com/questions/76563724}}.
The third challenge is mitigating response latency, where developers use asynchronous processing to prevent code execution from being blocked. For instance, issues arise when the ChatGPT API pauses the execution of the remaining code until it finishes processing\footnote{\url{https://stackoverflow.com/questions/75827468}}.

\begin{figure}[htbp] % [htbp] 表示图片位置的浮动参数    
\centering % 居中显示图片    
\includegraphics[width=0.45\textwidth]{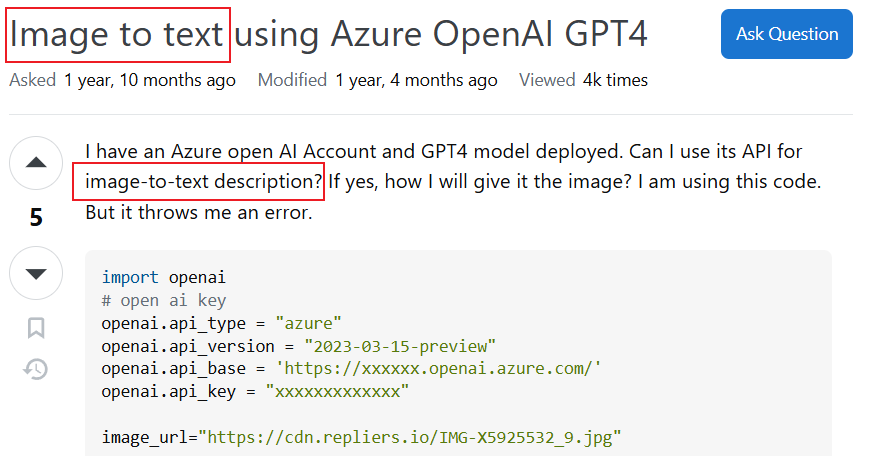}    
\caption{A sample post in the topic of multimodal processing.} % 图片标题    
\label{Fig:Multimodal_processing_Sample} % 图片标签，用于引用
\end{figure}

\textbf{A7: Multimodal Processing.}
With the growing demand for multimodal processing capabilities in chatbot conversations~\cite{zhang2024unveiling}, this topic focuses on multimodal processing, encompassing non-text data operations such as parsing PDFs, images, or CSV files. 
Developers frequently explore techniques for integrating diverse data formats with large language models, such as analyzing PDF files with the OpenAI API\footnote{\url{https://stackoverflow.com/questions/78511201}}, 
answering questions about CSV datasets through API calls\footnote{\url{https://stackoverflow.com/questions/78287134}}, 
and implementing image-to-text conversion using Azure OpenAI GPT 4 Vision\footnote{\url{https://stackoverflow.com/questions/76199709}}. As shown in Fig.~\ref{Fig:Multimodal_processing_Sample}, developers could not send images to GPT-4 since the service was not available then.

\textbf{A8: Framework and Toolchain Integration.}
This topic focuses on framework and toolchain challenges encountered during Chat API integration, primarily related to compatibility issues between third-party tools and development frameworks. A significant focus lies in \texttt{LangChain} compatibility, where developers report errors during chained operations, such as triggering \texttt{ValueError} exceptions when combining \texttt{ChatOpenAI} with \texttt{ChatPromptTemplate} via operators\footnote{\url{https://stackoverflow.com/questions/79250225}}. Deployment environment discrepancies further complicate implementations, as functions that work locally may fail in production environments, exemplified by \texttt{Axios} errors occurring exclusively in deployed web applications\footnote{\url{https://stackoverflow.com/questions/76627658}}. Furthermore, integration issues with third-party tools arise with extension startup failures\footnote{\url{https://stackoverflow.com/questions/79272471}}  and streaming response issues on platforms such as Visual Studio Code extensions or \texttt{Cloudflare} Workers\footnote{\url{https://stackoverflow.com/questions/77118020}}.

\textbf{A9: Custom Functionality.}
This topic focuses on the challenges faced in implementing custom functionality. A primary concern is plugin invocation, where developers seek to integrate ChatGPT plugins into their current development workflows\footnote{\url{https://stackoverflow.com/questions/77272952}}.
Another key aspect involves the control of prompts and responses. For example, by appending each new input to an existing prompt string rather than overwriting it\footnote{\url{https://stackoverflow.com/questions/69590991}}, developers ensure that the conversation history is preserved incrementally. Additionally, developers apply prompt engineering techniques to make ChatGPT generate single, direct responses instead of verbose narratives or off-topic elaborations\footnote{\url{https://stackoverflow.com/questions/76669635}}. This level of control reflects a broader trend toward fine-grained customization of model behavior to meet specific application requirements.

% 设置颜色、背景、边框等
\begin{tcolorbox}[
    colback=gray!10,    % 背景色
    colframe=black,   % 边框颜色
    boxrule=0.6pt,  
    sharp corners       % 直角边框
]
\textbf{Finding 3.}
44.2\% of the discussions focus on this topic, highlighting that the Chat API is the most important component within OpenAI’s API ecosystem. Based on the identified topics from the Chat API category, developers encounter several major challenges. These are primarily related to prompt design, including optimizing prompts to improve output quality and reduce usage costs, handling multimodal inputs, and addressing security concerns. In addition, developers face challenges such as compatibility issues caused by function or model deprecation, ensuring smooth conversation through asynchronous processing, and integrating with third-party tools.
\end{tcolorbox}

\subsection{Identified Topics for Embeddings API }

The Embeddings API enables developers to convert text into dense vector representations, facilitating semantic search, clustering, and integration with downstream applications. In our classification, this category accounts for 17.7\% of all developer discussions.
Through topic analysis, we identify seven distinct topics, which are introduced as follows.

\begin{figure}[htbp] % [htbp]   
\centering 
\includegraphics[width=0.45\textwidth]{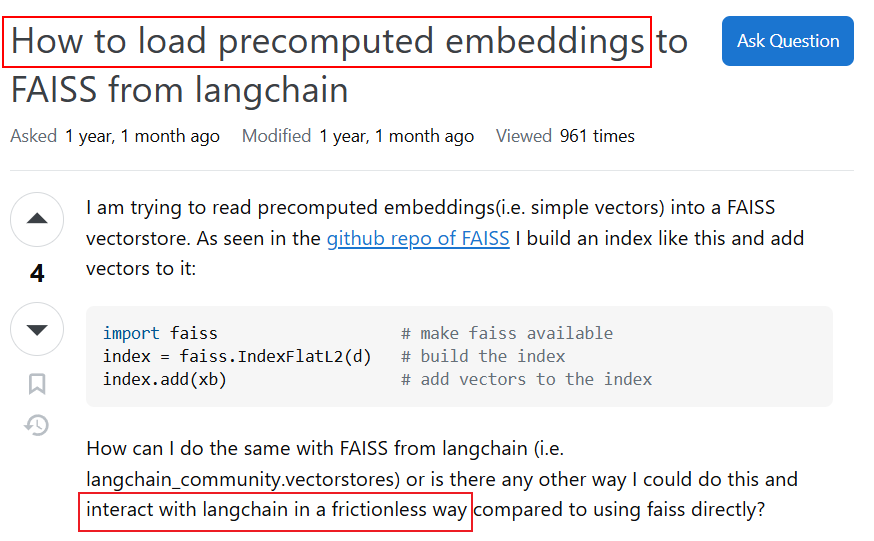}    
\caption{A sample post in the topic of vector database integration and maintenance.}     
\label{Fig:Vector_Database_Integration_and_Maintenance_Sample} 
\end{figure}

\textbf{B1: Vector Database Integration and Maintenance.}
This topic examines the operational challenges and compatibility issues associated with integrating and maintaining vector databases (e.g., \texttt{ChromaDB}) within embedding workflows. A key concern involves database management operations, such as adding embeddings to existing vector stores\footnote{\url{https://stackoverflow.com/questions/78465603}} and implementing safeguards to prevent the creation of duplicate embeddings when target directories already exist\footnote{\url{https://stackoverflow.com/questions/78462909}}. Additionally, compatibility challenges arise when integrating these databases with frameworks like \texttt{LangChain}. For instance, developers frequently encounter attribute errors when using the embedding function object with \texttt{ChromaDB}\footnote{\url{https://stackoverflow.com/questions/77004874}}, as well as difficulties loading precomputed embeddings into the Facebook AI Similarity Search (\texttt{FAISS}) vector store through \texttt{LangChain}'s interface\footnote{\url{https://stackoverflow.com/questions/77879936}}, as illustrated in Fig.~\ref{Fig:Vector_Database_Integration_and_Maintenance_Sample}.

\textbf{B2: API Request Failures and Rate Limitation.}
Developers frequently encounter challenges when working with OpenAI’s Embeddings API, such as encountering 429 errors when the number of requests exceeds the allowed limitation\footnote{\url{https://stackoverflow.com/questions/79341494}}. 
This highlights the importance of careful and efficient request management. Additional difficulties arise in scenarios involving large-scale data processing, such as generating embeddings from hundreds of CSV files one row at a time and uploading them to Pinecone using OpenAI embeddings\footnote{\url{https://stackoverflow.com/questions/78085597}}. In such a case, excessive batch requests may go beyond the permitted thresholds, resulting in failures or significant delays.

\begin{figure}[htbp]    
\centering 
\includegraphics[width=0.45\textwidth]{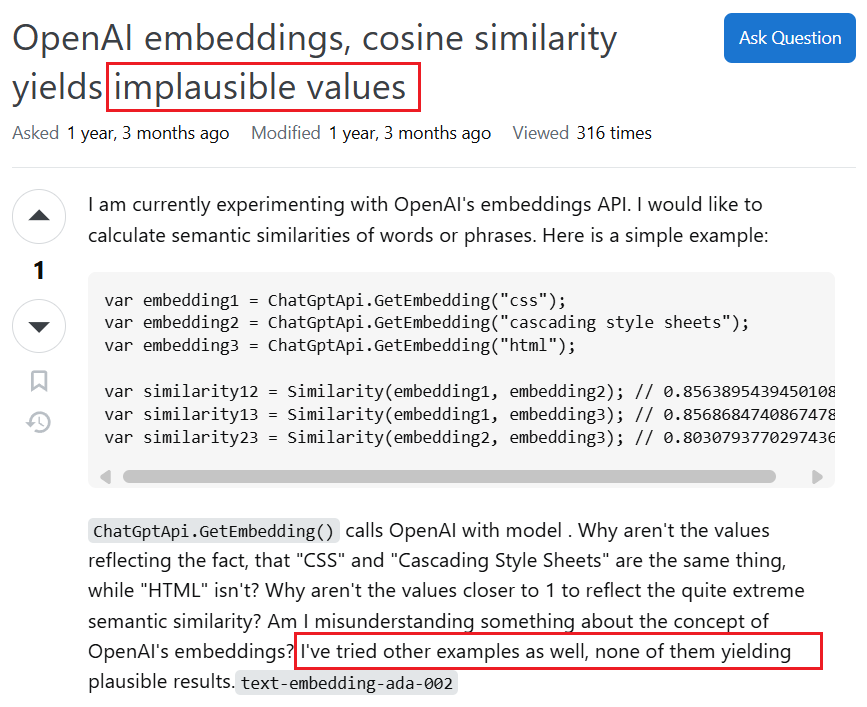}    
\caption{A sample post in the topic of attribute and query Errors.} 
\label{Fig:Attribute_and_Query_Errors_Sample} 
\end{figure}

\textbf{B3: Attribute and Query Errors.}
This topic primarily addresses issues stemming from incorrect usage of API methods and parameters, along with unexpected behavior in similarity search and query results. A common challenge involves attribute errors, such as attempts to access non-existent attributes like Embedding within the OpenAI module, which reflect misunderstandings or outdated usage of the API\footnote{\url{https://stackoverflow.com/questions/76511924}}. Additionally, developers report problems in similarity search scenarios. For example, queries executed with 
% LangChain Chroma 
\texttt{LangChain}'s integration with \texttt{Chroma} may fail to return relevant results\footnote{\url{https://stackoverflow.com/questions/79070763}}, and cosine similarity calculations may yield unexpected values when using OpenAI embeddings\footnote{\url{https://stackoverflow.com/questions/77607020}}, as illustrated in Fig.~\ref{Fig:Attribute_and_Query_Errors_Sample}.

\textbf{B4: Retrieval-Augmented Generation.}
Retrieval-Augmented Generation (RAG) integrates generative language models with external knowledge retrieval systems, utilizing embedding APIs to convert unstructured documents into semantic vectors for efficient similarity search and context-aware response generation. 
However, this integration encounters challenges. The first challenge involves the precision of document retrieval and the preservation of context. For example, developers often face difficulties in ensuring that the retrieved top $k$ documents directly contain relevant information\footnote{\url{https://stackoverflow.com/questions/78048414}}, which can result in inaccurate responses. Furthermore, the omission of source documents within processing chains\footnote{\url{https://stackoverflow.com/questions/77921898}} results in failures in maintaining contextual continuity, thereby compromising the reliability of generated answers.
Database storage limitations further contribute to these challenges. Constraints in storage capacity\footnote{\url{https://stackoverflow.com/questions/78748463}} hinder the scalability of knowledge bases, while inefficiencies in multi-document indexing
reveal difficulties in managing complex file hierarchies, such as the challenge of indexing 1,000 Python files across multiple subdirectories into a vector database and enabling RAG to leverage the folder structure\footnote{\url{https://stackoverflow.com/questions/79148668}}.

\textbf{B5: Embedding Model Configuration.}
This topic examines the challenges related to the configuration of embedding models, particularly concerning dimensionality and multimodal capabilities. A prevalent issue arises from discrepancies between embedding dimensions and model parameters, as demonstrated by errors indicating incorrect input vector formats\footnote{\url{https://stackoverflow.com/questions/75252902}} during cosine similarity searches and inconsistencies in dimensions\footnote{\url{https://stackoverflow.com/questions/78703704}} when using OpenAI embeddings in Python. Furthermore, the fusion of multimodal embeddings introduces distinct challenges, such as the combination of native image embeddings with text-based similarity searches\footnote{\url{https://stackoverflow.com/questions/79323765}}.

\textbf{B6: Data Management and Storage.}
This topic discusses the challenges related to document management and vector storage optimization when using the Embeddings API. For document management, developers often look for ways to verify whether a document already exists in a vector database, such as checking for document presence in a \texttt{Chroma} vector store using \texttt{LangChain}\footnote{\url{https://stackoverflow.com/questions/79340846}}. There is also significant interest in embedding structured data formats, including how to embed JSON documents using OpenAI's Embedding API\footnote{\url{https://stackoverflow.com/questions/78244309}}.
In terms of vector storage optimization, key issues include dealing with data redundancy, such as vector stores returning a large number of duplicate entries\footnote{\url{https://stackoverflow.com/questions/77555312}}, and designing efficient methods for detecting duplicates in vector databases\footnote{\url{https://stackoverflow.com/questions/76962785}}.

\textbf{B7: Advanced Applications.}
This topic explores advanced applications of the Embeddings API, focusing on challenges related to its sophisticated features, performance optimization, and complex usage scenarios. For instance, developers frequently seek to understand whether the \texttt{ConversationalRetrievalChain} retrieves answers directly from an in-memory vector database or invokes the OpenAI language model\footnote{\url{https://stackoverflow.com/questions/76813867}}. Another common issue involves compatibility limitations, such as errors triggered when certain vector encoding providers like \texttt{AzureOpenAI} are not supported by the \texttt{genai.vector.encode} function, resulting in messages like "Vector encoding provider AzureOpenAI is not supported"\footnote{\url{https://stackoverflow.com/questions/78505230}}. Moreover, developers often look for efficient strategies to save vectors after creating embeddings with \texttt{LangChain}\footnote{\url{https://stackoverflow.com/questions/77356114}}, aiming to avoid redundant reprocessing in future sessions.

% 设置颜色、背景、边框等
\begin{tcolorbox}[
    colback=gray!10,    % 背景色
    colframe=black,   % 边框颜色
    boxrule=0.6pt,  
    sharp corners       % 直角边框
]
\textbf{Finding 4.}
17.7\% of the discussions focus on this topic. Some discussions relate to vector database integration and optimization, while others focus on embedding techniques themselves, such as issues related to embedding dimensionality and multimodal fusion. As a key component of RAG systems, this area has also attracted considerable attention.
\end{tcolorbox}

\subsection{Identified Topics for Audio API}

OpenAI's Audio API offers both speech-to-text and text-to-speech capabilities.
 In our classification, this category accounts for 10.7\% of all developer discussions.
Through topic analysis, we identify six distinct topics, which are introduced as follows.

\begin{figure}[htbp] % [htbp] 表示图片位置的浮动参数    
\centering % 居中显示图片    
\includegraphics[width=0.45\textwidth]{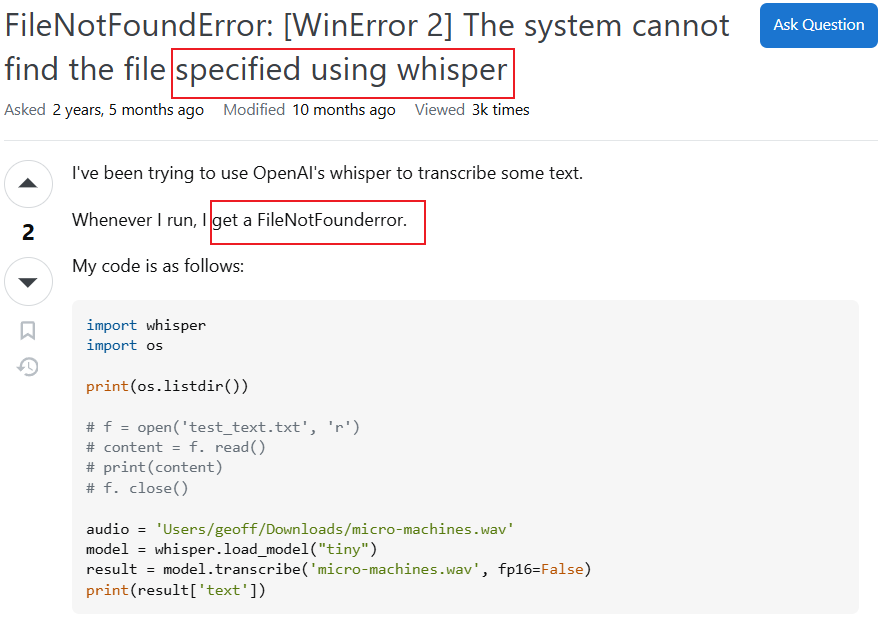}    
\caption{A sample post in the topic of model errors and file loading.} % 图片标题    
\label{Fig:Model_Errors_and_File_Loading_Sample} % 图片标签，用于引用
\end{figure}

\textbf{C1: Model Errors and File Loading.}
This topic focuses on runtime errors in the Whisper model and challenges related to loading audio files during API implementation. A major issue involves execution failures of the Whisper model, including initialization errors, inference interruptions, and parameter mismatches. For example, developers frequently encounter type errors\footnote{\url{https://stackoverflow.com/questions/75870435}} or attribute errors\footnote{\url{https://stackoverflow.com/questions/78172267}} related to the \texttt{transcribe} method. Additionally, issues with file path resolution and loading failures are common, with systems often misidentifying dependencies or failing to locate specified audio files, such as when Whisper cannot find the specified file\footnote{\url{https://stackoverflow.com/questions/73847516}}, as illustrated in Fig.~\ref{Fig:Model_Errors_and_File_Loading_Sample}.

\textbf{C2: Format Conversion and Stream Processing.}
This topic addresses challenges related to audio format transformations and real-time data stream integration in Audio API workflows. For instance, developers often seek solutions for converting arbitrary audio files into \texttt{np.ndarray} format when using OpenAI Whisper\footnote{\url{https://stackoverflow.com/questions/76779669}}. Additionally, there are efforts to integrate live audio streams, such as browser recordings or streaming media, with Whisper. Common scenarios include implementing live transcription by sending audio blobs to backend systems\footnote{\url{https://stackoverflow.com/questions/76947444}} or generating text-to-speech outputs from dynamically updating OpenAI data streams\footnote{\url{https://stackoverflow.com/questions/76085246}}.

\textbf{C3: Cross-Platform Deployment.}
This topic focuses on challenges in adapting OpenAI Whisper across diverse environments and hardware configurations. The first issue involves environment and framework compatibility, where developers encounter integration difficulties when deploying Whisper in non-standard environments, such as browsers, mobile frameworks (e.g., React Native), or packaging tools. For example, developers report failures in creating executable files with \texttt{PyInstaller} when OpenAI's Whisper is imported\footnote{\url{https://stackoverflow.com/questions/75594651}}. The second issue relates to GPU/CUDA resource utilization failures, where developers cannot leverage GPU acceleration due to missing NVIDIA drivers or improper library configurations, such as installing PyTorch without CUDA support\footnote{\url{https://stackoverflow.com/questions/75775272}}.

\textbf{C4: API Request Parameter Errors.}
When using the Audio API, developers often encounter issues with invalid API request parameters, such as the absence of the model parameter\footnote{\url{https://stackoverflow.com/questions/75594651}}, incorrect configuration of word timestamps\footnote{\url{https://stackoverflow.com/questions/76608484}}, or invalid API keys\footnote{\url{https://stackoverflow.com/questions/77505898}}. These issues can lead to errors, including the \texttt{InvalidRequestError}: Resource not found' exception.

\textbf{C5: Performance Optimization and Cost Management.}
This topic focuses on optimizing the performance of the Whisper model while managing associated usage costs. Developers often seek ways to obtain token usage metrics per minute from API responses\footnote{\url{https://stackoverflow.com/questions/75795242}}, suppress tokens during transcription\footnote{\url{https://stackoverflow.com/questions/76220528}}, and enhance the precision of multilingual speech recognition, particularly for technical programming terminology\footnote{\url{https://stackoverflow.com/questions/79304988}}. 
% To address these multilingual challenges, domain-specific fine-tuning with code-centric speech datasets and lexical augmentation techniques are employed to expand the tokenizer's coverage of specialized terminology~\cite{radford2023robust}.

\textbf{C6: Large-scale File Processing.}
This topic addresses challenges related to processing large audio and video files that exceed the Whisper API's default size limitations, with a focus on chunking strategies and post-processing alignment. Developers often face issues with splitting files into chunks while preserving transcription accuracy. For example, solutions are sought for transcribing large video files and implementing audio/video chunking workflows\footnote{\url{https://stackoverflow.com/questions/77058661}}. A key concern is synchronizing timestamp offsets across chunked audio transcripts to maintain temporal coherence in the final output\footnote{\url{https://stackoverflow.com/questions/76956604}}.

% 设置颜色、背景、边框等
\begin{tcolorbox}[
    colback=gray!10,    % 背景色
    colframe=black,   % 边框颜色
    boxrule=0.6pt,  
    sharp corners       % 直角边框
]
\textbf{Finding 5.}
10.7\% of the discussions focus on this topic, primarily related to OpenAI’s Whisper model and its use in speech-to-text and text-to-speech applications. Most of the discussions center on API parameter configuration, file processing, format transformation, and cross-platform deployment. Moreover, token cost optimization is also a frequently raised concern.
\end{tcolorbox}

\subsection{Identified Topics for Assistants API}

The Assistants API enables developers to easily build powerful AI assistants within their applications and supports improved function calling for third-party tools. 
In our classification, this category accounts for 10.0\% of all developer discussions.
Through topic analysis, we identify six distinct topics, which are introduced as follows.

\begin{figure}[htbp] % [htbp] 表示图片位置的浮动参数    
\centering % 居中显示图片    
\includegraphics[width=0.45\textwidth]{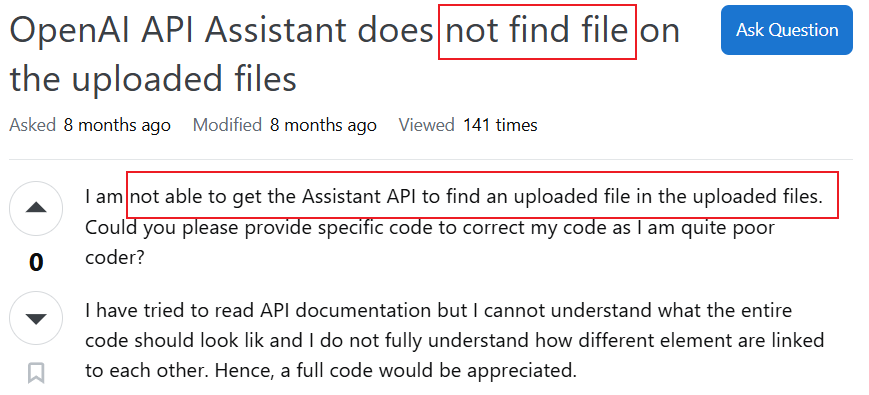}    
\caption{A sample post in the topic of core functions and fundamental configuration.} % 图片标题    
\label{Fig:Core_Functions_and_Fundamental_Configuration_Sample} % 图片标签，用于引用
\end{figure}

\textbf{D1: Core Functions and Fundamental Configuration.}
This topic focuses on the foundational capabilities and configurations of the Assistants API, covering three main aspects. First, core functionality implementation includes integrating tools such as multi-tool invocation, file search, and code interpreter setup. Developers often encounter issues, such as the Assistant being unable to locate uploaded files\footnote{\url{https://stackoverflow.com/questions/78713605}}, which reflects common challenges in file handling during API interactions (as shown in Fig.~\ref{Fig:Core_Functions_and_Fundamental_Configuration_Sample}). Second, error resolution and permission management involve API key validation, access control, and tool invocation failures. For example, some developers report receiving 404 errors when calling customized assistants through the ChatGPT API in R\footnote{\url{https://stackoverflow.com/questions/78437749}}, indicating typical configuration difficulties. Third, multi-source knowledge integration refers to the connection of various data sources, such as PDF documents and relational databases\footnote{\url{https://stackoverflow.com/questions/77058707}}, to the Assistants API. These aspects collectively represent the essential operations and configuration tasks required to effectively use the Assistants API.

\textbf{D2: Context Maintenance and Enhancement.}
This category focuses on memory management and contextual enhancement within the Assistants API framework, encompassing three key aspects. First, dialogue history preservation ensures conversation continuity across interactions. Developers often encounter issues, such as conflicts between different agent configurations that hinder the combined use of system messages and memory functions\footnote{\url{https://stackoverflow.com/questions/77954484}}. Second, custom prompt engineering aims to refine assistant behavior by using tailored templates that enforce specific output formats. Third, external knowledge integration involves combining the API with retrieval-augmented generation methods\footnote{\url{https://stackoverflow.com/questions/78331627}} or specialized tools, such as the \texttt{Tavily} API\footnote{\url{https://stackoverflow.com/questions/78334865}}, to improve context retrieval and response relevance.

\textbf{D3: Tool Integration and Model Extension.}
This topic explores tool integration and model extension within the Assistants API ecosystem, focusing on three key areas. First, tool interoperability and model compatibility involve connecting external frameworks such as \texttt{LangChain}\footnote{\url{https://stackoverflow.com/questions/78392233}}, enabling collaborative workflows across multiple AI models. Second, file parsing and data stream processing highlight challenges in managing structured formats like JSON and CSV files\footnote{\url{https://stackoverflow.com/questions/78755971}}. Third, expanding the application scope includes advanced use cases such as performing database operations—including create, read, update, and delete tasks—and integrating visualizations. For instance, developers have attempted to execute SQL queries through the Assistants API to support more dynamic data interactions\footnote{\url{https://stackoverflow.com/questions/78414172}}.

\textbf{D4: Real-time Response and Stream Processing.}
This category highlights real-time response mechanisms and stream optimization within the Assistants API. Developers often build Flask-based chatbots to support live interactions through efficient data streaming\footnote{\url{https://stackoverflow.com/questions/78761075}}. Integration with multi-platform systems (such as WebSocket protocols and Next.js applications using the AI SDK) is also explored to improve real-time user experiences across web environments\footnote{\url{https://stackoverflow.com/questions/78702505}}. In addition, developers face challenges when processing large documents and managing response format limitations, particularly in scenarios that require balancing file size with performance efficiency\footnote{\url{https://stackoverflow.com/questions/77941413}}.

\textbf{D5: Threads and Automation Processing.}
This topic explores thread management and automation workflows within the Assistants API. The first focus is on data sharing and thread efficiency, where developers aim to optimize multi-session interactions by reusing context within threads, as seen in strategies for sharing context using the same assistant\footnote{\url{https://stackoverflow.com/questions/78215919}}. However, prolonged use of a single thread may lead to high token consumption, due to the 100,000-message limit per thread\footnote{\url{https://platform.openai.com/docs/assistants}}, as historical messages accumulate and increase token usage over time. The second focus involves automating complex workflows (such as machine learning pipelines or data visualization tasks)  by using natural language commands, allowing developers to streamline processes and reduce manual coding efforts\footnote{\url{https://stackoverflow.com/questions/78326755}}.

\textbf{D6: Function Calling.}
This topic addresses the challenges of invoking custom functions using the OpenAI Assistants API. A common issue involves parameter handling and JSON format processing during function execution. For instance, invalid request errors often arise from incorrect parameter transmission\footnote{\url{https://stackoverflow.com/questions/76661527}}. Developers also encounter difficulties with asynchronous operations and potential memory leaks. These include problems with asynchronous function calls\footnote{\url{https://stackoverflow.com/questions/77670252}} and cases where server-based Flask applications using the Assistants API terminate unexpectedly, possibly due to memory-related issues\footnote{\url{https://stackoverflow.com/questions/77699248}}.

% 设置颜色、背景、边框等
\begin{tcolorbox}[
    colback=gray!10,    % 背景色
    colframe=black,   % 边框颜色
    boxrule=0.6pt,  
    sharp corners       % 直角边框
]
\textbf{Finding 6.}
10.0\% of the discussions focus on this topic. This type of API mainly assists developers in building powerful AI assistants. As a result, most of the issues are related to tool integration, such as challenges in effectively using these APIs (e.g., operation tasks and configuration tasks) and issues related to threat management. In addition, there are also discussions on prompt engineering and the latest LLM techniques (such as RAG) to better support assistant development.
\end{tcolorbox}

\subsection{Identified Topics for Fine-tuning API}

The Fine-tuning API allows developers to tailor pre-trained models to specific tasks or domains using custom datasets, enhancing their adaptability and performance. 
In our classification, this category accounts for 8.4\% of all developer discussions.
Through topic analysis, we identify six distinct topics, which are introduced as follows.

\textbf{E1: Basic Error Handling.}
This topic addresses foundational technical challenges and error handling during the fine-tuning process. Key issues include dependency conflicts caused by framework errors in PyTorch or Hugging Face, as well as hardware limitations such as GPU memory constraints. Model loading failures are common, often due to corrupted weight files, format mismatches, or incorrect storage paths when saving results. API request exceptions arise from version deprecation and parameter configuration errors. For instance, developers may encounter operational failures, such as the "That model does not exist" error when executing \texttt{CLI} commands for GPT-3 fine-tuned models, typically caused by outdated API version dependencies\footnote{\url{https://stackoverflow.com/questions/77699248}}.

\textbf{E2: Customization via Model Fine-tuning.}
This topic focuses on model customization and functional adaptation through the fine-tuning API, enabling tailored model behaviors and optimized training strategies. A key aspect is controlling model outputs to meet domain-specific requirements, such as ensuring ChatGPT responds solely based on fine-tuned data\footnote{\url{https://stackoverflow.com/questions/76976251}} or adapting it for email conversation analysis\footnote{\url{https://stackoverflow.com/questions/75783524}}. These applications highlight methods to narrow response scope and improve task-specific performance. Additionally, researchers explore parameter-efficient fine-tuning (PEFT) techniques, such as comparing adapter tuning and prefix tuning\footnote{\url{https://stackoverflow.com/questions/74710732}}, to enhance training efficiency and model adaptability.

\textbf{E3: Dataset Construction and Output Control.}
This topic explores the challenges of structuring datasets and ensuring output consistency when using OpenAI's Fine-tuning API. A primary concern is data formatting and validation, where users seek guidance on creating compliant datasets. For example, discussions emphasize the importance of structuring datasets for GPT-3 fine-tuning\footnote{\url{https://stackoverflow.com/questions/70531364}} and ensuring adherence to the JSONL format, with each JSON object separated by a newline character, such as verifying that each line represents a valid JSON object\footnote{\url{https://stackoverflow.com/questions/75935259}}. Another critical aspect is output consistency and validation, addressing issues related to aligning model outputs with predefined patterns or custom data. Examples include inquiries into ensuring responses are solely based on fine-tuned datasets and resolving misclassifications in \texttt{gpt-3.5-turbo} fine-tuned models, where outputs deviate from defined classes\footnote{\url{https://stackoverflow.com/questions/77847649}}.

\begin{figure}[htbp] % [htbp] 表示图片位置的浮动参数    
\centering % 居中显示图片    
\includegraphics[width=0.45\textwidth]{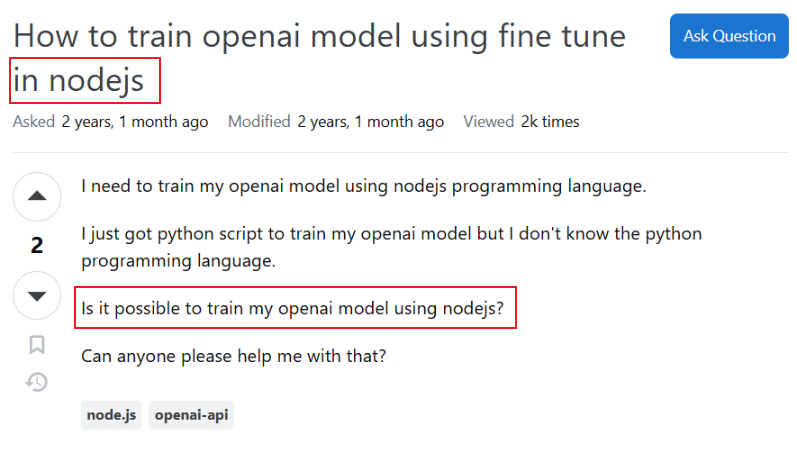}    
\caption{A sample post in the topic of API basic usage.} % 图片标题    
\label{Fig:API_Basic_Usage_Sample} % 图片标签，用于引用
\end{figure}

\textbf{E4: API Basic Usage.}
This topic addresses challenges developers face when implementing OpenAI's Fine-tuning API, with a focus on API invocation failures and environment-specific integration issues. A common problem involves API call failures, often caused by improper data formatting or invalid authentication. Developers also encounter terminal command recognition errors, such as the message "command not found: openai," typically due to incorrect environment variable configurations or missing dependencies\footnote{\url{https://stackoverflow.com/questions/75343008}}. Additionally, functional anomalies can occur after fine-tuning, such as the loss of tool call capabilities in customized GPT-4 mini models\footnote{\url{https://stackoverflow.com/questions/79086015}}. Another key challenge is environmental integration, where developers attempt to invoke the Fine-tuning API in specific environments, such as using Node.js for fine-tuning OpenAI models\footnote{\url{https://stackoverflow.com/questions/75469378}} to align with project architectures (as shown in Fig.~\ref{Fig:API_Basic_Usage_Sample}).

\textbf{E5: Model Management and Resource Optimization.}
This topic explores challenges and strategies related to managing customized models and optimizing resource usage within the Fine-tuning API. For instance, developers must verify whether a fine-tuned model has been successfully deleted\footnote{\url{https://stackoverflow.com/questions/78716179}}, as errors may occur if the model does not exist\footnote{\url{https://stackoverflow.com/questions/77111087}}. To minimize operational costs, developers often implement on-demand deployment strategies, temporarily activating fine-tuned models for inference tasks and undeploying them during idle periods to avoid incurring hourly computational charges\footnote{\url{https://stackoverflow.com/questions/79192524}}.

% 设置颜色、背景、边框等
\begin{tcolorbox}[
    colback=gray!10,    % 背景色
    colframe=black,   % 边框颜色
    boxrule=0.6pt,  
    sharp corners       % 直角边框
]
\textbf{Finding 7.}
8.4\% of the discussions focus on this topic. These APIs mainly aim to adapt a pre-trained model to a specific task by continuing its training on domain-specific datasets. As a result, most discussions center around the choice of training datasets, API usage, fine-tuning methods (such as PEFT), issues related to JSON format, and the management and optimization of fine-tuned models.
\end{tcolorbox}

\subsection{Identified Topics for Image Generation API}

The Image Generation API enables developers to generate and edit images.
In our classification, this category accounts for 2.5\% of all developer discussions.
Through topic analysis, we identify three distinct topics, which are introduced as follows.

\textbf{F1: Image Input Formats and API Configuration.}
This topic addresses common challenges encountered when working with OpenAI's Image Generation API, particularly those related to image format specifications and API configuration. Developers frequently face issues arising from three main aspects. 
First, input image format errors occur when images submitted to the edit or variation endpoints fail to meet format requirements. For example, using an RGBA format instead of the required RGB format\footnote{\url{https://stackoverflow.com/questions/74773173}}. Second, API configuration and initialization failures often result from improper module imports or incorrect usage of configuration classes during setup. 
Third, request parameter errors emerge when API calls to the generation or edit endpoints contain missing or malformed parameters.

\textbf{F2: API Usage Limits and Cross-Environment Invocation Conflicts.}
This topic explores errors resulting from API access restrictions and environment-specific compatibility issues. Developers often face request rejections due to exceeded API quotas or unconfigured billing settings, reflecting limitations at the account level. For instance, data capacity constraints, such as image uploads exceeding the 4 MB limit\footnote{\url{https://stackoverflow.com/questions/78030548}}, can lead to failed requests. Moreover, cross-environment compatibility issues may arise when using the API across different frameworks (e.g., JavaScript), where discrepancies in parameter validation can cause errors, such as a missing image property in requests to the variations endpoint\footnote{\url{https://stackoverflow.com/questions/77322543}}.

\textbf{F3: Generated Image Processing and Version Updates.}
This category highlights key challenges related to image persistence, operational workflows, and SDK compatibility. One common issue involves retrieving images generated by OpenAI and saving them to external storage services such as Amazon S3. This task requires developers to handle transient, URL-based outputs and integrate them with third-party cloud storage solutions\footnote{\url{https://stackoverflow.com/questions/74729716}}. It is essential to account for the expiration of ephemeral URLs and implement reliable storage mechanisms.
Additionally, developers encounter difficulties in managing historical records of generated images and executing batch operations across multiple image outputs\footnote{\url{https://stackoverflow.com/questions/78589430}}. Furthermore, SDK version updates can lead to compatibility issues, where deprecated attributes or methods disrupt existing API integrations. These challenges necessitate careful version management and the adoption of appropriate code migration strategies\footnote{\url{https://stackoverflow.com/questions/78177250}}.

% 设置颜色、背景、边框等
\begin{tcolorbox}[
    colback=gray!10,    % 背景色
    colframe=black,   % 边框颜色
    boxrule=0.6pt,  
    sharp corners       % 直角边框
]
\textbf{Finding 8.}
2.5\% of the discussions focus on this topic, primarily related to generating and editing images. Common issues mainly focus on image format and API configuration. In addition, some discussions raise concerns about limitations and compatibility issues when using this type of API.
\end{tcolorbox}

\subsection{Identified Topics for Code Generation API}

OpenAI's Code Generation API (such as Codex and GPT-4.1) translates natural language instructions into executable code, supporting multiple programming languages. It is widely used in scenarios like automated development, data analysis, and building AI assistants. 
In our classification, this category accounts for 1.7\% of all developer discussions.
Through topic analysis, we identify three distinct topics, which are introduced as follows.

\begin{figure}[htbp] % [htbp] 表示图片位置的浮动参数    
\centering % 居中显示图片    
\includegraphics[width=0.45\textwidth]{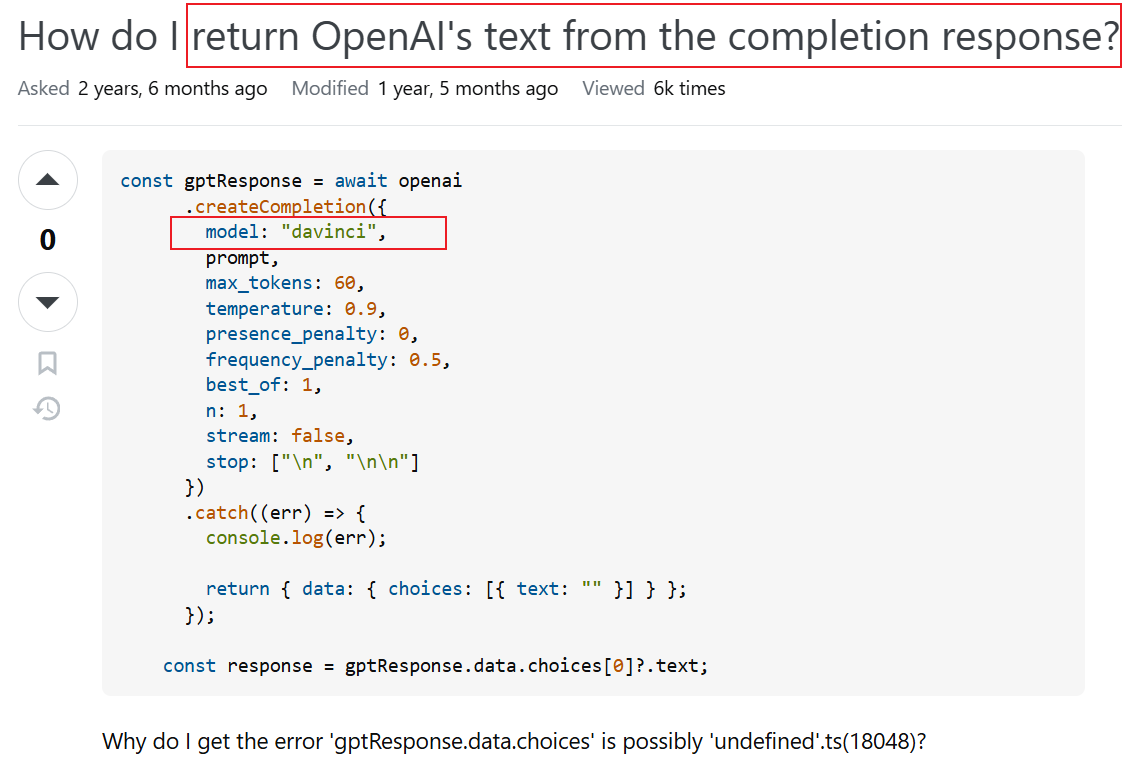}    
\caption{A sample post in the topic of fundamental API usage and integration issues.} % 图片标题    
\label{Fig:Fundamental_API_Usage_and_Integration_Issues_Sample} % 图片标签，用于引用
\end{figure}

\textbf{G1: Fundamental API Usage and Integration Issues.}
This topic covers common challenges in basic API usage and tool compatibility. The first issue involves errors in calling API functions, such as incorrect parameter settings, environment setup failures, or mishandling responses. For example, Fig.~\ref{Fig:Fundamental_API_Usage_and_Integration_Issues_Sample} shows that developers sometimes have trouble extracting text from API responses\footnote{\url{https://stackoverflow.com/questions/76193116}}. This happens because of incorrect parameter settings or disabled streaming, which causes incomplete outputs.
The second issue relates to compatibility problems with specific programming languages (e.g., PHP, C++) or development tools (e.g., \texttt{Monaco} Editor). A common case is variable passing errors during stream completion tasks\footnote{\url{https://stackoverflow.com/questions/78177250}}.

\textbf{G2: Code Generation.}
This topic focuses on natural language-to-code generation and prompt engineering techniques to control code output. One key application is converting natural language queries into SQL (Structured Query Language) syntax for database interactions. Developers use prefix and suffix prompt structures in Codex models to improve code generation accuracy. For example, analyzing different prompt configurations shows how the context affects the output\footnote{\url{https://stackoverflow.com/questions/73094271}}. Additionally, strategies to remove unwanted response prefixes in Codex models help prevent unnecessary explanatory text before generated JSON arrays\footnote{\url{https://stackoverflow.com/questions/75564196}}.

\textbf{G3: Model Parameters and Output Control.}
This topic covers three key aspects of parameter optimization and output control. First, model parameter configuration and sampling control involve adjusting hyperparameters, such as the relationship between the \texttt{best\_of} parameter and nucleus sampling\footnote{\url{https://stackoverflow.com/questions/72947447}}. These techniques help generate diverse candidates (nucleus sampling) and select the best value for the parameter \texttt{best\_of}. Developers also want to know how to use the Codex API to get embeddings for a given code snippet\footnote{\url{https://stackoverflow.com/questions/72986749}}.
Second, context management parameters require careful calibration of the context window length. The combined token count of input prompts and \texttt{max\_tokens} outputs should stay within model-specific limits\footnote{\url{https://stackoverflow.com/questions/75403409}}.
The third aspect focuses on output format control and managing intermediate processes, such as removing leading spaces in shell command outputs\footnote{\url{https://stackoverflow.com/questions/79354638}} or retrieving intermediate reasoning steps when token limits are exceeded\footnote{\url{https://stackoverflow.com/questions/76164731}}.

% 设置颜色、背景、边框等
\begin{tcolorbox}[
    colback=gray!10,    % 背景色
    colframe=black,   % 边框颜色
    boxrule=0.6pt,  
    sharp corners       % 直角边框
]
\textbf{Finding 9.}
1.7\% of the discussions focus on this topic, which involves APIs designed to generate code based on natural language descriptions. The main discussions include issues related to API usage, such as parameter settings, environment setup, and handling of responses. Additionally, some discussions focus on prompt design and post-processing of the generated code to further improve code quality.
\end{tcolorbox}

\subsection{Identified Topics for GPT Actions API}

The GPT Actions API allows developers to extend models' functionality through custom API integrations and multi-operation workflows. 
In our classification, this category accounts for 1.4\% of all developer discussions.
Through topic analysis, we identify two distinct topics, which are introduced as follows.

\textbf{H1: Custom Action and Multi-Operation Management.}
This topic focuses on developing and coordinating customized actions within GPT systems to enhance ChatGPT's functionality by integrating external services. The main focus is on creating and linking custom API interactions, allowing natural language inputs to be converted into parameterized API requests. For example, a custom ChatGPT action could be set up to retrieve real-time financial data (such as the latest stock price) from the Alpha Vantage API using the OpenAI Schema\footnote{\url{https://stackoverflow.com/questions/78939365}}. The secondary focus is on expanding domain-specific capabilities by managing multiple operations. A typical example is creating several operations for the same domain using GPT Builder.

\textbf{H2: External Data Connectivity and File Processing.}
This topic focuses on using GPT Actions API to connect external data sources and manage file operations through RESTful APIs. Key aspects include integrating real-time data for dynamic content, such as news and weather, and parsing structured documents like PDFs or Google Docs. It also covers the need for accessing databases or cloud storage services. The final aspect involves configuring API requests to support file uploads, such as adjusting the OpenAPI schema to send files along with textual data\footnote{\url{https://stackoverflow.com/questions/77498087}}.

% 设置颜色、背景、边框等
\begin{tcolorbox}[
    colback=gray!10,    % 背景色
    colframe=black,   % 边框颜色
    boxrule=0.6pt,  
    sharp corners       % 直角边框
]
\textbf{Finding 10.}
1.4\% of the discussions focus on this topic, which involves APIs that can perform real-world tasks by integrating with external tools. The main focus is on challenges related to combining with external tools, such as converting natural language inputs into parameterized API requests and connecting with external data sources.
\end{tcolorbox}

\subsection{Identified Topics for Others}

This category encompasses edge cases involving OpenAI API functionalities beyond the eight types, including deprecated or niche services such as the \texttt{Moderation} API, \texttt{Batch} API, and \texttt{Classification} API, which are less discussed. 
In our classification, this category accounts for 3.4\% of all developer discussions.
Through topic analysis, we identify two distinct topics, which are introduced as follows.

\textbf{I1: General Technical Barriers.}
This topic does not focus on specific OpenAI API types but addresses common technical challenges encountered during development, such as foundational integration, environment configuration, and proxy or network connectivity issues. For example, a missing \texttt{openai} module in an application can prevent basic API initialization\footnote{\url{https://stackoverflow.com/questions/77528940}}.

\textbf{I2: OpenAI API Edge Cases.}
This topic covers other OpenAI APIs that do not fall under the eight primary API categories, such as the \texttt{Moderation} API and deprecated APIs. These APIs are discussed less frequently in developer forums. For instance, "How to see if OpenAI (Node.js) createModeration response 'flagged' is true" demonstrates the use of the \texttt{Moderation} API to detect flagged content\footnote{\url{https://stackoverflow.com/questions/74836812}}. Additionally, deprecated APIs, such as the \texttt{Classification} API, have been discontinued since December 2022. The \texttt{Answers} API has also been integrated into the \texttt{Responses} API's document search module.

% 设置颜色、背景、边框等
\begin{tcolorbox}[
    colback=gray!10,    % 背景色
    colframe=black,   % 边框颜色
    boxrule=0.6pt,  
    sharp corners       % 直角边框
]
\textbf{Finding 11.}
3.4\% of the discussions focus on this topic, primarily including edge cases unrelated to the eight previously mentioned API categories. For example, it covers issues related to technical barriers and less frequently discussed APIs, such as the Moderation API.
\end{tcolorbox}

\section{Discussions}
\label{sec:discussions}

In this section, we first present the implications for stakeholders related to OpenAI APIs. We then analyze potential threats to the validity of our empirical study.

\subsection{Implications}

Our empirical study identified the challenges developers face when using OpenAI APIs. 
Based on the analyzed trends, difficulties, and identified topics, we provide practical implications for developers, LLM vendors, and researchers. 
Our key implications can be summarized as follows:

\textbf{Creating comprehensive tutorials and documentation.}
Some issues with the OpenAI API stem from developers' insufficient understanding of fundamental LLM concepts (such as prompt engineering~\cite{marvin2023prompt}, the embedding concept, fine-tuning, and RAG), particularly among those without relevant backgrounds. In such cases, existing tutorials and documentation often fail to comprehensively cover the concepts and technical details involved in LLM application development. 
To better support non-expert LLM developers, there is an urgent need for LLM vendors to provide clearer and more comprehensive tutorials and documentation, as suggested by previous studies~\cite{morovati2024common,chen2025empirical}. 
These resources should offer detailed guidance on core LLM concepts, best practices, and common pitfalls, while minimizing misunderstandings and barriers during development.

\textbf{Enhancing Version Compatibility and API Deprecation Management.}
Among the identified challenges, version compatibility issues (e.g., SDK upgrades in the Chat API) and API deprecation (e.g., model-loading failures due to the deprecated Fine-tuning API) are widespread when using OpenAI APIs.
To address frequent SDK and API updates, LLM vendors should establish robust version management mechanisms to enhance system stability and user experience, including allowing users to select specific model versions to avoid compatibility risks from automatic upgrades, and publishing clear deprecation policies with advance notice of relevant timelines.
For API deprecation issues, LLM vendors should provide transparent deprecation processes to enhance service stability and user trust, including early announcements, reasonable transition periods, detailed migration guides, and compatibility support.

\textbf{Improving Context Management in Conversations.} 
Maintaining context information across multiple rounds of dialogue remains a major challenge, especially for chat-based APIs (such as the Chat API and Assistants API). Notice that this issue is mainly limited to these types of APIs, as other OpenAI APIs (such as Image Generation API and Embeddings API) do not require maintaining contextual information.
In recent years, developers have commonly faced limitations related to context length and dialogue continuity. As the number of conversation turns increases, large language models are prone to issues such as truncation or topic drift~\cite{liang2024c5}. To address this challenge, researchers should explore more effective interpolation and extrapolation methods~\cite{pawar2024and} to improve LLMs' ability to manage conversational context. In addition, LLM vendors need to further expand context window sizes or implement more efficient built-in memory management strategies~\cite{yin2024llm}, thereby enabling more coherent and natural multi-turn interactions.

\textbf{Optimizing OpenAI API Cost Management.}
Our empirical study shows that the cost of using OpenAI APIs (such as Chat API and Audio API) is a major concern for developers. To optimize these costs, developers should first optimize input prompts and limit output length to reduce unnecessary token consumption. Secondly, developers can implement local caching mechanisms to reduce the cost of redundant API calls. Additionally, developers should set usage limits for their API keys to avoid exceeding their cost budget. Finally, regularly monitoring token usage through the OpenAI console can help developers adjust their strategies, reducing token usage while maintaining service quality.

\textbf{Constructing implicit API usage knowledge base.}
Currently, OpenAI's official documentation primarily focuses on the functional descriptions and basic usage of the APIs. However, during actual development and deployment, developers often need to follow a series of implicit best practices to avoid issues such as program crashes, training anomalies, or performance degradation (such as Audio API, Fine-tuning API, and Image Generation API). This API usage knowledge is typically scattered across community discussions, technical blogs, and developer forums, and has yet to be systematically compiled. 
To address this issue, researchers should construct a comprehensive OpenAI API usage knowledge base by mining community knowledge~\cite{zhong2009mapo}. Specifically, researchers should first search Stack Overflow for questions and answers related to OpenAI API usage, extracting common patterns in error handling, performance optimization, and best practices; secondly, researchers should analyze open-source projects on GitHub, examining project issues, pull requests, and code changes before and after modifications, to gather development insights from real-world projects; finally, researchers should mine technical blogs to collect firsthand experiences and optimization tips shared by developers.

\textbf{Developing code quality assurance tools.}
When developing LLM-related code, researchers should explore several directions and develop corresponding tools to improve code quality.
For example, researchers could develop static analysis tools to automatically detect deprecated OpenAI APIs~\cite{haryono2021characterization,wang2025llms} and API misuse~\cite{yang2024demystifying}. This would allow automated identification and reporting of issues in the code, reducing the burden of manual review.
In addition, researchers should develop API recommendation tools~\cite{wei2022api} that intelligently suggest appropriate APIs and parameter settings based on developers' needs and the code context, providing more accurate API usage guidance and improving development efficiency.
Finally, researchers should build tools to automatically identify bugs in the code, analyze their symptoms and root causes, and provide repair suggestions~\cite{jia2021symptoms}, helping developers locate and fix problems more quickly and enhancing the stability and reliability of the code.

\subsection{Threats to Validity}
%参考那些API挑战的论文去写
%要全面
In this subsection, we discuss potential threats to our empirical study and corresponding alleviation strategies.

\textbf{Internal Threats.}
The first threat concerns the identification of OpenAI API-related tags. To ensure quality, we followed the approach of prior studies~\cite{process2,significance-3}, selecting relevant tags based on their significance and relevance.
% 增加了in the data annotation的主观性
The second threat involves the subjectivity in manual inspection during data collection and in data annotation.
To alleviate this threat, the second author and the third author independently conducted manual inspections. The inter-rater reliability, measured by Cohen’s Kappa, indicates almost perfect agreement between annotators.
The third threat pertains to the determination of the optimal number of topics and the naming of topics. To identify the optimal number of topics, we adopted optimization approaches, such as varying the number of topics from 2 to 20 (in increments of 1) and calculating their corresponding coherence scores.
To improve the quality of topic naming, we followed previous studies~\cite{chen2020comprehensive, lda1} and utilized an open card sort method~\cite{fincher2005making}.

\textbf{External Threats.}
The external validity of our study raises concerns about the generalizability of our results.
The first threat is that our empirical study only analyzes the challenges developers face when using the OpenAI API. However, OpenAI's models, such as the GPT series, are among the most widely adopted across the industry, serving as the backbone for applications like chatbots, code generation, and writing assistance. Moreover, OpenAI provides stable and mature API services, allowing developers to easily access and switch between different model versions. Therefore, the challenges we identified, such as the complexities of prompt engineering, token-based cost management, non-deterministic outputs, and operation as black boxes, are broadly applicable and practically significant. In future work, we plan to extend our analysis to include APIs offered by other vendors such as DeepSeek, DeepMind, and Meta.
The second threat is that OpenAI continuously updates its APIs, which may introduce new challenges over time. Our empirical study only analyzes posts up until January 2025, and ongoing monitoring of more recent posts will be necessary to capture newly emerging challenges.
The third threat is that we only considered posts from the Stack Overflow forum. However, Stack Overflow hosts a large and active global community of developers and experts and maintains high standards for content quality. In future work, we plan to include discussions from LLM vendors' official developer forums and gather relevant codebases and issues from GitHub projects to further enhance and validate our findings.

\textbf{Construct Threats.}
The primary construct threat concerns our analysis of popularity and difficulty.
To alleviate this threat, for popularity, we follow prior studies~\cite{alshangiti2019developing, shah2025towards} and measure it using the annual user counts and the number of related questions.
For difficulty, we also follow prior work
~\cite{significance-1}, considering two main aspects: (1) the proportion of posts with an accepted answer, and (2) the time taken for a post to receive an accepted answer.

\section{Related Work}
\label{sec:related}

APIs are a fundamental component of modern software development. However, developers often encounter various challenges in their practical usage. As a result, identifying and understanding these challenges has become a prominent research focus in the field of software engineering. Researchers commonly analyze discussions on Q\&A platforms (such as Stack Overflow) and issues in GitHub to identify these challenges.
For example, Scoccia et al.~\cite{process1}
investigated the common difficulties developers encounter when building desktop web applications using frameworks like Electron and NW.js.
Venkatesh et al.~\cite{venkatesh2016client}  identified the primary concerns of client developers when using Web APIs, identifying five dominant topics per API that cover at least 50\% of related questions and highlighting persistent issues that API providers should address.
Rosen et al.~\cite{rosen2016mobile} investigated the primary challenges mobile developers face, revealing that their most common questions cover app distribution, mobile APIs, data management, sensors and context, mobile tools, and user interface development. 
Beddiar et al.~\cite{beddiar2020classification} presented a supervised learning approach to classify Stack Overflow posts related to Android API issues, aiming to assist developers in identifying and addressing common challenges in Android development.

With the rise of machine learning and deep learning, researchers have become interested in the challenges faced in the development of such software.
For example, Islam et al.~\cite{islam2019developers}
presented a large-scale analysis on popular machine learning libraries, such as TensorFlow, Keras, scikit-learn, and Weka, to identify the most common challenges developers face across different stages of the machine learning pipeline.
Alshangiti et al.~\cite{alshangiti2019developing} identified the primary challenges developers face in machine learning application development, revealing that most difficulties arise during data preprocessing and model deployment phases, often due to a lack of implementation knowledge and limited expert support within the community.

Different from previous studies, we are the first to analyze the challenges developers face when using the OpenAI API.
To achieve this, we collected 2,874 OpenAI API-related posts from Stack Overflow and categorized them into nine distinct API types.
We then applied topic modeling techniques to identify the specific challenges associated with each API category.
Based on our findings, we propose a set of implications for developers, LLM vendors, and researchers. These include providing more comprehensive tutorials and documentation (especially for prompt engineering), enhancing API compatibility and managing deprecated APIs effectively, optimizing API usage cost management, and developing tools for API search and misuse detection tailored to OpenAI APIs.

\section{Conclusion}
\label{sec:conclusion}
% Conclusion

This study provides the first comprehensive empirical analysis of OpenAI API-related developer discussions on Stack Overflow, revealing both the popularity trends and distinct technical challenges across nine major API categories. By leveraging topic modeling and manual categorization of 2,874 posts, we uncover key difficulties such as non-deterministic outputs, prompt engineering complexity, token-based cost management, and integration hurdles with multimodal inputs and third-party tools. These insights not only highlight the evolving nature of API usage in the era of large language models but also offer actionable implications for developers, LLM providers, and researchers to improve usability, documentation, and API design.

\section*{Acknowledgements}
 
This research was partially supported by the National Natural Science Foundation of China (Grant No. 61202006), the Open Project of State Key Laboratory for Novel Software Technology at Nanjing University (Grant No. KFKT2024B21), and the Postgraduate Research \& Practice Innovation Program of Jiangsu Province (Grant No. SJCX24\_2022).

\bibliographystyle{IEEEtran}
\bibliography{main}

\end{document}